\documentclass[aps,prl,onecolumn,showpacs,preprintnumbers]{revtex4}
\usepackage{eurosym}
\usepackage{amsmath}
\usepackage{dcolumn}
\usepackage{bm}
\usepackage{subfigure}
\usepackage{amsfonts}
\usepackage{amssymb}
\usepackage{makeidx}
\usepackage{epsfig}
\usepackage{graphicx}

\begin{document}

\title{Macroscopic irreversibility and decay to kinetic equilibrium\\
for classical hard-sphere systems}
\author{M. Tessarotto}
\affiliation{Department of Mathematics and Geosciences, University of Trieste, Via
Valerio 12, 34127 Trieste, Italy}
\affiliation{Institute of Physics and Research Center for Theoretical Physics and
Astrophysics, Faculty of Philosophy and Science, Silesian University in
Opava, Bezru\v{c}ovo n\'{a}m.13, CZ-74601 Opava, Czech Republic}
\author{C. Cremaschini}
\affiliation{Institute of Physics, Faculty of Philosophy and Science, Silesian University
in Opava, Bezru\v{c}ovo n\'{a}m.13, CZ-74601 Opava, Czech Republic}
\date{\today }

\begin{abstract}
In this paper the conditions are investigated for the occurrence of the
so-called macroscopic irreversibility property and the related phenomenon of
decay to kinetic equilibrium which may characterize the $1-$body probability
density function (PDF) associated with hard-sphere systems. The problem is
set in the framework of the axiomatic "ab initio" approach to classical
statistical mechanics recently developed [Tessarotto \textit{et al}.,
2013-2017] and the related establishment of an exact kinetic equation
realized by Master equation for the same kinetic PDF. As shown in the paper
the task involves the introduction of a suitable functional of the $1-$body
PDF here identified with the \textit{Master kinetic information}. The goal
is to show that, provided the same PDF is realized in terms of an arbitrary
suitably-smooth particular solution of the Master kinetic equation the two
properties indicated above are indeed realized and that the same functional
is unrelated either with the Boltzmann-Shannon entropy and the Fisher
information.
\end{abstract}

\maketitle

\bigskip

%\frontmatter

%\email{cremasch@sissa.it}

%\email{M.Tessarotto@cmfd.univ.trieste.it}

\section{1 - Introduction}

In this investigation the problem is posed of the \emph{proof-of-principle}
for two phenomena which characterize the statistical description of $N-$body
hard-sphere systems and laying at the very foundations of classical
statistical mechanics (CSM) and kinetic theory alike. The issue, more
precisely, is related to the physical conditions for the possible occurrence
of the so-called \emph{property of macroscopic irreversibility }(PMI) and
the consequent one represented by the \emph{decay to kinetic equilibrium}
(DKE).

In the following the case is considered of the so-called Boltzmann-Sinai
classical dynamical system (CDS) \cite{noi1} which advances in time the
microscopic state $\mathbf{x}\equiv \left\{ \mathbf{x}_{1},...,\mathbf{x}%
_{N}\right\} $ of a set of $N$ extended like particles represented by smooth
hard spheres \cite{noi7} of diameter $\sigma >0$, with $\mathbf{x}_{i}\equiv
\left( \mathbf{r}_{i},\mathbf{v}_{i}\right) $, $\mathbf{r}_{i}$ and $\mathbf{%
v}_{i}$ denoting Newtonian center of mass state, position and velocity of
the $i-$th particle. The same particles are assumed: A) subject to
instantaneous (unary, binary and multiple) elastic collisions which leave
unchanged the particles angular momenta and B) immersed in a bounded domain $%
\Omega $ of the Euclidean space $%
%TCIMACRO{\U{211d} }%
%BeginExpansion
\mathbb{R}
%EndExpansion
^{3}$ of finite measure.

For definiteness, the treatment is set in the framework of the "\emph{ab
initio}" axiomatic approach to CSM recently-developed in Refs.\cite%
{noi1,noi2,noi3,noi4,noi5,noi6,noi7} (see also Refs.\cite{noi8,noi9,noi10})
and the consequent establishment of an exact, i.e., non-asymptotic, kinetic
equation \cite{noi3}, denoted as Master kinetic equation. \ The new approach
radically departs from standard approaches to be found in the literature
such as the Boltzmann and Enskog kinetic equations \cite%
{Boltzmann1972,Enskog,CHAPMA-COWLING} which apply only in an asymptotic
sense for large $N-.$body hard sphere systems, i.e., in which the number of
particles $N$ is considered $\gg 1$. \ In fact, the remarkable
distinguishing feature of the new equation is that, unlike the
aforementioned kinetic equations, it holds in the case of the \emph{finite
Boltzmann-Sinai CDS} (shortly referred to as $S_{N}-$CDS), namely for
arbitrary hard-sphere systems having a \emph{finite number} $N$ of particles%
\textbf{\ }and in which each particle is allowed to have, in addition, a
\emph{finite-size, }namely is characterized by a finite diameter $\sigma >0,$
and a\emph{\ }finite-mass $m>0.$ \textbf{\ }

The goal of the paper is to pose in such a context the problem of the
existence of both PMI and DKE holding in the case of finite hard-sphere
systems.  The conjecture is that - just as the ergodicity property of the $%
S_{N}-$CDS \cite{Sinai1970,Sinai1989} - the possible occurrence of such
phenomena in actual physical, i.e. necessarily finite, systems, might/should
not depend on the number $N$ of constituent particles of the system. In
particular we intend to show that these properties actually emerge as
necessary implications of the Master kinetic equation itself. Incidentally,
in doing so, the finiteness requirement on the $S_{N}-$CDS completely rules
out for further possible consideration either the Boltzmann or the Enskog
kinetic equations, these equations being manifestly inapplicable to the
treatment of systems of this type.

Specifically, in the following the case $N>2$\ is considered everywhere,
which is by far the most physically-relevant one. In this occurrence, in
fact, non-trivial $2-$body occupation coefficients arise (see related
notations which are applicable for $N>2$ \ recalled in Appendices A and B
below). For completeness the case $N=2$\ is nevertheless briefly discussed
in Appendix D.

\subsection{1A - Motivations and background}

Both properties indicated in the title concern the statistical behavior of
an ensemble $S_{N}$ of like particles which are advanced in time by a
suitable $N-$body classical dynamical system, here identified with the $%
S_{N}-$CDS. Specifically they arise in the context of the kinetic
description of the same CDS, i.e., in terms of the corresponding $1-$body
(kinetic) probability density function (PDF) $\rho _{1}^{(N)}(t)\equiv \rho
_{1}^{(N)}(\mathbf{x}_{1},t).$ The latter is required to belong to a
suitable functional class $\left\{ \rho _{1}^{(N)}(\mathbf{x}_{1},t)\right\}
$ of smooth and strictly positive ordinary functions which are particular
solutions of the relevant kinetic equation.

In fact, PMI should be realized by means of a suitable, but still possibly
non-unique, functional which should be globally defined in the future (i.e.,
for all times $t\geq t_{o}$ being $t_{o}$ a suitable initial time) bounded
and non-negative and therefore to be identified with the notion of
information measure. Most importantly, however, the same functional, to be
referred to here as \emph{Master kinetic information }(MKI), should also
exhibit a continuously-differentiable and monotonic, i.e., in particular
decreasing, time-dependence.

Regarding, instead, the second property of DKE this concerns\emph{\ }the
asymptotic behavior of the $1-$body PDF $\rho _{1}^{(N)}(t)\equiv \rho
_{1}^{(N)}(\mathbf{x}_{1},t)$ which, accordingly, should be globally defined
and decay for $t\rightarrow +\infty $ to a stationary and spatially-uniform
Maxwellian PDF
\begin{equation}
\rho _{1M}^{(N)}(\mathbf{v}_{1})=\frac{n_{o}}{\pi ^{3/2}\left(
2T_{o}/m\right) ^{3/2}}\exp \left\{ -\frac{m\left( \mathbf{v}_{1}-\mathbf{V}%
_{o}\right) ^{2}}{2T_{o}}\right\} ,  \label{MAXWELLIAN-PDF}
\end{equation}%
where $\left\{ n_{o}>0,T_{o}>0,\mathbf{V}_{o}\right\} $\emph{\ }are suitable
constant fluid fields.

Both PMI and DKE correspond to physical phenomena which might/should
possibly arise in disparate classical $N-$body systems. The clue is
represented by the ubiquitous occurrence of kinetic equilibria and
consequently, in principle, also of the corresponding possible manifestation
of macroscopic irreversibility and\emph{\ }decay processes. Examples of the
former ones are in principle easy to be found, ranging from neutral fluids
\cite{ikt1,ikt2,ikt4,ikt7,ikt8} to collisional/collisionless and
non-relativistic/relativistic gases and plasmas \cite%
{Cr2010,Cr2011,Cr2011a,Crema2012,Crema2013,Crema2013a,Crema2013b,Crema2014}.
However, the most notable example is provided by dilute gases characterized
by a large number of particles ($N\equiv \frac{1}{\varepsilon }\gg 1$) and a
small (i.e., infinitesimal) diameter $\sigma \sim O(\varepsilon ^{1/2})$ of
the hard-spheres. In fact, the property of macroscopic irreversibility
indicated above is related to the Carnot's second Law of Classical
Thermodynamics and the historical attempt of its first-principle-proof
performed originally by Ludwig Boltzmann in 1872 \cite{Boltzmann1972}.
Indeed both phenomena lie at the very root of Boltzmann and Grad kinetic
theories \cite{Boltzmann1972,Grad}, although a different characterization of
the concept of PMI is actually involved. In particular, the goal set by
Boltzmann himself in his 1872 paper was the proof of Carnot's Law providing
at the same time also a possible identification of thermodynamic entropy.
This was achieved in terms of what is nowadays known as Boltzmann-Shannon
(BS) statistical entropy, which is identified with the phase-space moment
\begin{equation}
M_{X_{E}}(\rho _{1}(t))\equiv \int\limits_{\Gamma _{1}}d\mathbf{x}_{1}\rho
_{1}(\mathbf{x}_{1},t)X_{E}(\mathbf{x}_{1},t)=-\int\limits_{\Gamma _{1}}d%
\mathbf{x}_{1}\rho _{1}(\mathbf{x}_{1},t)\ln \frac{\rho _{1}(\mathbf{x}%
_{1},t)}{A_{1}}\equiv S(\rho _{1}(t)).  \label{BS-entropy}
\end{equation}%
Here $X_{E}(\mathbf{x}_{1},t)\equiv -\ln \frac{\rho _{1}(\mathbf{x}_{1},t)}{%
A_{1}},$ $\rho _{1}(\mathbf{x}_{1},t)$ and $A_{1}$ denote respectively the
BS entropy density, an arbitrary particular solution of the Boltzmann
equation for which the same phase-space integral exists and an arbitrary
positive constant. In fact, according to the Boltzmann H-theorem the same
functional should satisfy the so-called \emph{entropic inequality}
\begin{equation}
\frac{\partial }{\partial t}S(\rho _{1}(t))\geq 0
\label{Entropic enequality}
\end{equation}%
while, furthermore, the \emph{entropic equality condition}%
\begin{equation}
\frac{\partial }{\partial t}S(\rho _{1}(t))=0\Leftrightarrow \rho _{1}(%
\mathbf{x}_{1},t)=\rho _{1M}^{(N)}(\mathbf{v}_{1})  \label{entropic equality}
\end{equation}%
should hold. The latter equation implies therefore that, if $\rho _{1}(t)$
and $S(\rho _{1}(t))$ exist globally, then necessarily $lim_{t\rightarrow
+\infty }\rho _{1}(\mathbf{x}_{1},t)=\rho _{1M}^{(N)}(\mathbf{v}_{1})$.
However, both Boltzmann and Grad theories are actually specialized to the
treatment of the so-called \emph{Boltzmann-Grad limit }obtained introducing,
first, the dilute-gas ordering $\sigma \sim O(\varepsilon ^{1/2})$ with $%
\varepsilon \equiv \frac{1}{N}\ll 1$ and, then, taking the continuum limit $%
\varepsilon \rightarrow 0$ (for a review of the topic see again Ref.\/\cite%
{noi10}).

Nevertheless, the possible realization of either PMI or DKE depends
critically on the prescription of the functional class $\left\{ \rho ^{(N)}(%
\mathbf{x}_{1},t)\right\} ,$ so that their occurrence is actually
non-mandatory. Indeed, both cannot occur - also for Boltzmann and Grad
kinetic theories \cite{noi6} - if the $N-$body (microscopic) probability
density function $\rho ^{(N)}(\mathbf{x,}t)$ is identified with the
deterministic $N-$body PDF \cite{noi1}, namely the $N-$body phase-space
Dirac delta. This is defined as $\delta (\mathbf{x-x}(t))\equiv
\prod\limits_{1=1,N}\delta (\mathbf{x}_{i}\mathbf{-x}_{i}(t)),$ with $%
\mathbf{x}\equiv \left\{ \mathbf{x}_{1},...,\mathbf{x}_{N}\right\} $
denoting the state of the $N-$body system and $\mathbf{x}(t)\equiv \left\{
\mathbf{x}_{1}(t),...,\mathbf{x}_{N}(t)\right\} $ is the image of an
arbitrary initial state $\mathbf{x}(t_{o})\equiv \mathbf{x}_{o}$ generated
by the same $N-$body CDS. \ That such a PDF necessarily must realize an
admissible particular solution of the $N-$body Liouville equation follows,
in fact, as a straightforward consequence of the axioms of classical
statistical mechanics \cite{noi1}.

Despite these premises, however, the case of a finite Boltzmann-Sinai CDS,
which is characterized by a finite number of particles $N$ and/or a
finite-size of the hard spheres and/or a dense or locally-dense system, is
more subtle and - as explained below - even unprecedented since it has
actually remained unsolved to date. The reasons are that:

\begin{itemize}
\item First, Boltzmann and Grad kinetic theories are inapplicable to the
finite Boltzmann-Sinai CDS.

\item Second, as shown in Ref.\cite{noi6} the Boltzmann-Shannon entropy
associated with an arbitrary particular solution $\rho ^{(N)}(t)\equiv \rho
^{(N)}(\mathbf{x}_{1},t)$ of the Master kinetic equation, i.e., the
functional $S(\rho _{1}^{(N)}(t))\equiv M_{X_{E}}(\rho _{1}^{(N)}(t)),$ in
contrast to $S(\rho _{1}(t))\equiv M_{X_{E}}(\rho _{1}(t)),$ is exactly
conserved in the sense that identically%
\begin{equation}
\frac{\partial S(\rho _{1}^{(N)}(t))}{\partial t}\equiv 0
\end{equation}

must hold. As a consequence the validity itself of Boltzmann H-theorem
breaks down in the case of the Master kinetic equation.

\item Third, an additional motivation is provided by the conjecture that
both PMI and DKE might occur only if the Boltzmann-Grad limit is actually
performed, i.e., only in validity of Boltzmann equation and H-theorem.
\end{itemize}

Hence the question which arises is whether in the case of a finite
Boltzmann-Sinai CDS the phenomenon of DKE may still arise. Strong
indications seem to be hinting at such a possibility. In this regard the
example-case which refers to the statistical description of an
incompressible viscous Navier-Stokes granular fluid in terms of the Master
kinetic equation is relevant and suggests that this may be indeed the case.
In fact, as shown in Ref. \cite{noi8}, in such a case the decay of the fluid
velocity field occurring in a bounded domain necessarily requires the
existence of DKE too. However, besides the construction of the kinetic
equation appropriate for such a case, a further unsolved issues lies in the
determination of the functional class $\left\{ \rho _{1}^{(N)}(\mathbf{x}%
_{1},t)\right\} $ for which both PMI and DKE should/might be realized. In
particular, the possible occurrence of both PMI and DKE should correspond to
suitably-smooth, but nonetheless still arbitrary, initial conditions $%
\left\{ \rho _{1}^{(N)}(\mathbf{x}_{1},t_{o})\right\} $. These should
warrant that in the limit $t\rightarrow +\infty ,$ $\rho _{1}^{(N)}(\mathbf{x%
}_{1},t)$ uniformly converges to the spatially-homogeneous and stationary
Maxwellian PDF $\rho _{1M}(\mathbf{v}_{1})$ (\ref{MAXWELLIAN-PDF}). Such a
result, however, is highly non-trivial since it should rely on the
establishment of a global existence theorem for the same $1-$body PDF $\rho
_{1}^{(N)}(\mathbf{x}_{1},t)$ - namely holding in the whole time axis $%
I\equiv
%TCIMACRO{\U{211d} }%
%BeginExpansion
\mathbb{R}
%EndExpansion
$, besides the same $1-$body phase space $\Gamma _{1}$ - for the involved
kinetic equation which is associated with the $S_{N}-$CDS. In the context of
the Boltzmann equation in particular, despite almost-endless efforts this
task has actually not been accomplished yet, the obstacle being
intrinsically related to the asymptotic nature of the Boltzmann equation
\cite{noi7}. In fact for the same equation it is not known\ in satisfactory
generality whether smooth enough solutions of the same equation exist which
satisfy the $H-$theorem inequality and decay asymptotically to kinetic
equilibrium \cite{Cercignani1982,Villani}.

\subsection{1B - Goals of the investigation}

In a series of papers \cite{noi1,noi2,noi3,noi4,noi5,noi6,noi7} a new
kinetic equation has been established for hard sphere systems subject to
elastic instantaneous collisions, denoted as Master kinetic equation (see
Appendix A). Its remarkable feature is that unlike the Boltzmann and Enskog
kinetic equations \cite{Boltzmann1972,Enskog} the new kinetic equation and
its corresponding Master collision operator are exact, i.e., they hold for
an arbitrary \emph{finite} $N-$body hard-sphere system $S_{N}$. In other
words this means that in such a context $S_{N}$ is allowed to have in
principle an arbitrary \emph{constant} and \emph{finite number} ($N$) of%
\emph{\ }hard spheres, each one characterized by a \emph{finite }diameter $%
\sigma >0$ and a \emph{finite} mass $m>0$.

These peculiar features follow uniquely as a consequence of the new approach
to classical statistical mechanics developed in Refs.\cite{noi1,noi2,noi3}
and referred to as "\textit{ab initio}" axiomatic approach. As shown in the
same references (for a review see also Ref.\cite{noi9}), this is based on
the adoption of appropriate \emph{extended functional setting} and
physics-based \emph{modified collision boundary conditions} (MCBC; see
Appendix B) \cite{noi1,noi2} which are prescribed in order to advance in
time across arbitrary (unary, binary or multiple) collision events the $N-$%
body PDF. The related physical interpretation is intuitive. It can be
viewed, in fact, as the jump condition for the $N-$body PDF along the
phase-space Lagrangian trajectory $\left\{ \mathbf{x}(t)\right\} $ for an
ensemble of $N$ tracer particles \cite{noi0,noi9} following the same
deterministic trajectory and undergoing a collision event at a suitable
collision time.

Based on the discovery of the Master kinetic equation, a host of new
developments have opened up. These concern in particular the investigation
of the conceptual aspects and implications of the same equation which
include (for an extended discussion see also Refs.\cite{noi4,noi5,noi6,noi7}%
):

\begin{enumerate}
\item \emph{The determination of the Master H-theorem: }as pointed out in
Refs.\cite{noi5,noi6} based on the discovery of a family of generalized
collisional invariants, the Master kinetic equation is found to admit a
constant H-theorem in terms of the Boltzmann-Shannon entropy $S_{1}(\rho
_{1}^{(N)}(t))$.

\item \emph{The derivation of the Boltzmann kinetic equation in terms of the
Master kinetic equation}. The Boltzmann equation can be recovered in an
asymptotic sense when the so-called \emph{dilute-gas asymptotic ordering} is
introduced in the Master kinetic equation (see Refs.\cite{noi7,noi10}).

\item \emph{The global validity of the Master kinetic equation: }the Master
kinetic equation has been shown to hold globally in time \cite{noi7}.
\end{enumerate}

However, the question arises of the possible occurrence of both PMI and DKE
for arbitrary finite-size and/or dense systems of hard spheres. The
example-case recently pointed out \cite{noi8}, corresponding to the
statistical description of an incompressible Navier-Stokes granular fluid,
suggests that this may be indeed the case. The goal of the present paper is
to propose a new approach, referred to as \emph{PMI/DKE theory}, to the
treatment of PMI and DKE for hard-sphere systems described by means of the
Master kinetic equation. The core of the new theory is the first-principle
proof of both microscopic irreversibility and DKE properties holding for the
Master kinetic equation.

For this purpose, first, in Section 2, the MKI functional is explicitly
determined. We display in particular its construction method (see \emph{%
No.\#1- \#4 MKI Prescriptions)}. Second, in Section 2, based on the theory
of the Master kinetic equation earlier developed \cite{noi3} and suitable
integral and differential identities (see Appendices A and B), the
properties of the MKI functional are investigated. These concern in
particular the establishment of appropriate inequalities holding for the
same functional \ (THM.1, subsection 2A), the signature of the time
derivative of the same functional (THM.2, subsection 2B) and the property of
DKE holding for a suitable class of $1-$body PDFs (THM.3, subsection 2C). In
the subsequent sections 3 and 4, the issue of the consistency of the
phenomena of PMI and DKE with microscopic dynamics is posed together with
the physical interpretation and implications of the theory. The\ goal is to
investigate the relationship of the DKE-theory developed here with the
microscopic reversibility principle and\ the Poincar\'{e} recurrence
theorem. Finally in Section 5 the conclusions of the paper are drawn and
possible applications/developments of the theory are pointed out.

\section{2 - Axiomatic prescriptions for the MKI functional}

In view of the considerations given above we now proceed constructing an
explicit possible realization of the MKI functional in terms of suitable
axiomatic prescriptions. \textbf{This }should be intended as a functional $%
I_{M}\left( \rho _{1}^{(N)}(t)\right) $ of the $1-$body PDF $\rho
_{1}^{(N)}(t),$\ with $\rho _{1}^{(N)}(t)\equiv \rho _{1}^{(N)}(\mathbf{x}%
_{1},t)$ being identified with an arbitrary particular solution of the
Master kinetic equation (see Eq.(\ref{App-1}) in Appendix A). The same PDF
is assumed globally defined, a property which in view of Ref.\cite{noi7} is
warranted in particular if the initial PDF $\rho _{1o}^{(N)}(\mathbf{x}_{1})$
(prescribed by the initial problem (\ref{App-1}) at the initial time $%
t_{o}\in I$) belongs to the functional class of stochastic $1-$body PDFs,
i.e., strictly positive ordinary functions $\rho _{1}^{(N)}(t)\equiv \rho
_{1}^{(N)}(\mathbf{x}_{1},t)$ which are smoothly-differentiable. Hereon for
definiteness the case $N>2$\ is considered (see Appendix D for comments
about the treatment of the case $N=2$).

More specifically, first (\emph{MKI Prescription No.\#0 }), the functional
class $\left\{ \rho _{1}^{(N)}(\mathbf{x}_{1},t)\right\} $ is prescribed.
This is identified with the subset of stochastic particular solutions of the
Master kinetic equation which are images of the corresponding initial PDFs $%
\left\{ \rho _{1o}^{(N)}(\mathbf{x}_{1})\right\} .$ In turn $\left\{ \rho
_{1o}^{(N)}(\mathbf{x}_{1})\right\} $ is the ensemble of all initial PDFs
for which the functional $I_{M}\left( \rho _{1o}^{(N)}(\mathbf{x}%
_{1})\right) $ exists. In the remainder $\left\{ \rho _{1}^{(N)}(\mathbf{x}%
_{1},t)\right\} $ will be referred to as \emph{functional class of the
admissible stochastic PDFs}.

Second (\emph{MKI Prescription No.\#1 }), the functional $I_{M}(\rho
_{1}^{(N)}(t))$ should be suitably prescribed so that, assuming that by
construction the initial value $I_{M}\left( \rho _{1o}^{(N)}(\mathbf{x}%
_{1})\right) $ exists, then the same functional necessarily must exist
globally in the future, i.e., for all $t\geq t_{o}$ where $t_{o}\in I$ is a
suitable initial time$.$ Third, we shall require (\emph{MKI Prescription
No.\#2}) $I_{M}(\rho _{1}^{(N)}(t))$ to be real, non-negative and bounded in
$\left\{ \rho _{1}^{(N)}(\mathbf{x}_{1},t)\right\} $ in the sense%
\begin{equation}
0\leq I_{M}(\rho _{1}^{(N)}(t))\leq 1  \label{MKI-1}
\end{equation}%
so that it can be interpreted as an information measure associated with the $%
1-$body PDF $\rho _{1}^{(N)}(t)\equiv \rho _{1}^{(N)}(\mathbf{x}_{1},t).$
For this reason the previous inequalities will be referred to as\ \emph{%
information-measure inequalities}. Fourth, for consistency with the property
of macroscopic irreversibility (PMI), (\emph{MKI Prescription No.\#3}) $%
I_{M}(\rho _{1}^{(N)}(t))$ is prescribed in terms of a smoothly
time-differentiable and monotonically time-decreasing functional in the
sense that in the same time-subset the inequality:
\begin{equation}
\frac{\partial }{\partial t}I_{M}(\rho _{1}^{(N)}(t))\leq 0  \label{MKI-2}
\end{equation}%
should identically apply $\forall t\geq t_{o},$ so that by construction
\begin{equation}
0\leq I_{M}(\rho _{1}^{(N)}(t))\leq I_{M}\left( \rho _{1o}^{(N)}(\mathbf{x}%
_{1})\right) \leq 1.  \label{MKI-2bis}
\end{equation}%
which implies that $I_{M}(\rho _{1}^{(N)}(t))$ is also globally defined for
all $t\in I\equiv
%TCIMACRO{\U{211d} }%
%BeginExpansion
\mathbb{R}
%EndExpansion
$ with $t\gtrsim t_{o}.$ In addition, if $\left. \frac{\partial }{\partial t}%
I_{M}(\rho _{1}^{(N)}(t))\right\vert _{t=t_{o}}\neq 0$, without loss of
generality its initial value $I_{M}\left( \rho _{1o}^{(N)}(\mathbf{x}%
_{1})\right) $ can always be set such that%
\begin{equation}
I_{M}\left( \rho _{1o}^{(N)}(\mathbf{x}_{1})\right) =1.  \label{MKI-2ter}
\end{equation}%
As a fifth condition, in order to warrant the existence of DKE we shall
require (\emph{MKI Prescription No.\#4 }) the functional $I_{M}(\rho
_{1}^{(N)}(t))$ to be prescribed in such a way that at an arbitrary time $%
t\in I,$ with $t\gtrsim t_{o},$ the vanishing of both $I_{M}(\rho
_{1}^{(N)}(t))$ and its time derivative $\frac{\partial }{\partial t}%
I_{M}(\rho _{1}^{(N)}(t))$ should occur if and only if the $1-$body PDF
solution of the Master kinetic equation coincides with kinetic equilibrium.
As a consequence, for the functional $I_{M}(\rho _{1}^{(N)}(t))$ the
following propositions should be equivalent%
\begin{equation}
\left\{
\begin{array}{c}
I_{M}(\rho _{1}^{(N)}(t))=0 \\
\frac{\partial }{\partial t}I_{M}(\rho _{1}^{(N)}(t))=0%
\end{array}%
\Leftrightarrow \rho _{1}^{(N)}(\mathbf{x}_{1},t)=\rho _{1M}^{(N)}(\mathbf{v}%
_{1}),\right.  \label{MKI-4a}
\end{equation}%
with $\rho _{1M}^{(N)}(\mathbf{v}_{1})$ being a kinetic equilibrium PDF of
the form (\ref{MAXWELLIAN-PDF}).

The implication of \emph{MKI Prescriptions \#0-\#4 }is that, provided a
realization of the MKI can be found in the functional class of the initial
conditions indicated above $\left\{ \rho _{1o}^{(N)}(\mathbf{x}_{1})\right\}
$ the existence of both PMI and DKE for the Master kinetic equation would
actually be established.

In the sequel the goal is to show that the MKI functional can be identified
with the functional%
\begin{equation}
\left\{
\begin{array}{c}
I_{M}(\rho _{1}^{(N)}(t),\mathbf{b})\equiv \frac{K_{M}(\rho _{1}^{(N)}(t),%
\mathbf{b})}{K_{Mo}}, \\
K_{M}(\rho _{1}^{(N)}(t),\mathbf{b})=-\int\limits_{\Gamma _{1(1)}}d\mathbf{x}%
_{1}\overline{\Theta }_{1}^{(\partial \Omega )}(\overline{\mathbf{r}})M(%
\mathbf{v}_{1},\mathbf{b})\frac{\rho _{1}^{(N)}(\mathbf{x}_{1},t)}{\widehat{%
\rho }_{1}^{(N)}(\mathbf{x}_{1},t)}\frac{\partial ^{2}\widehat{\rho }%
_{1}^{(N)}(\mathbf{x}_{1},t)}{\partial \mathbf{r}_{1}\cdot \partial \mathbf{r%
}_{1}}, \\
K_{Mo}=\sup \left\{ 1,K_{M}(\rho _{1o}^{(N)}(\mathbf{x}_{1}),\mathbf{b}%
)\right\} .%
\end{array}%
\right.  \label{MKI-functional-1}
\end{equation}%
Here $\rho _{1}^{(N)}(t)\equiv \rho _{1}^{(N)}(\mathbf{x}_{1},t),$ $\rho
_{1o}^{(N)}(\mathbf{x}_{1})$ and $\widehat{\rho }_{1}^{(N)}(t)\equiv
\widehat{\rho }_{1}^{(N)}(\mathbf{x}_{1},t)$ are respectively the $1-$body
PDF solution of the initial problem associated with the Master kinetic
equation (see Eq.(\ref{App-1}) in Appendix A), with $\rho _{1o}^{(N)}(%
\mathbf{x}_{1})$ being the initial PDF, and the \emph{renormalized }$1-$body
PDF
\begin{equation}
\widehat{\rho }_{1}^{(N)}(\mathbf{x}_{1},t)\equiv \frac{\rho _{1}^{(N)}(%
\mathbf{x}_{1},t)}{k_{1}^{(N)}(\mathbf{r}_{1},t)},  \label{App-00}
\end{equation}%
with $k_{1}^{(N)}(\mathbf{r}_{1},t)$ being the $1-$body occupation
coefficient whose definition is recalled in Appendix B (see Eq.(\ref{App-4}%
)). As a consequence in the previous equation $\frac{\rho _{1}^{(N)}(\mathbf{%
x}_{1},t)}{\widehat{\rho }_{1}^{(N)}(\mathbf{x}_{1},t)}=k_{1}^{(N)}(\mathbf{r%
}_{1},t)$. Furthermore $\overline{\Theta }_{1}^{(\partial \Omega )}(%
\overline{\mathbf{r}})$ is the boundary theta function given by Eq.(\ref%
{boundary theta function}) (see Appendix A) and finally $\frac{1}{2}M(%
\mathbf{v}_{1},\mathbf{b})$ denotes the \emph{directional kinetic energy}
along $\mathbf{b}$ carried by particle $1\mathbf{,}$ namely the dynamical
variable%
\begin{equation}
M(\mathbf{v}_{1},\mathbf{b})\equiv \left( \mathbf{v}_{1}\cdot \mathbf{b}%
\right) ^{2},  \label{MOMENT-1}
\end{equation}%
with $\mathbf{b}$ denoting an arbitrary constant unit vector. Hence
\begin{equation}
M(\mathbf{v}_{1},\mathbf{v}_{2},\mathbf{b})=\frac{1}{2}\left[ M(\mathbf{v}%
_{1},\mathbf{b})+M(\mathbf{v}_{2},\mathbf{b})\right]
\label{TOTAL DIRECTIONAL ENERGY}
\end{equation}%
identifies the corresponding \emph{total directional kinetic energy} carried
by particles $1$ and $2$. As a consequence it follows that if $K_{M}(\rho
_{1o}^{(N)}(\mathbf{x}_{1}),\mathbf{b})\geq 1$ or $0\leq K_{M}(\rho
_{1o}^{(N)}(\mathbf{x}_{1}),\mathbf{b})<1$, then consistent with (\ref%
{MKI-2bis}) by construction respectively one should obtain%
\begin{equation}
I_{M}(\rho _{1o}^{(N)},\mathbf{b})=\left\{
\begin{array}{c}
1, \\
K_{M}(\rho _{1o}^{(N)}(\mathbf{x}_{1}),\mathbf{b}).%
\end{array}%
\right.  \label{MKI-6}
\end{equation}

\subsection{2A - Proof of the non-negativity of the MKI information measure}

The strategy adopted for the proof of the \emph{No.\#1 }and \emph{No.\#2 MKI
Prescriptions} is to prove initially the validity of the information-measure
left inequality in Eq.(\ref{MKI-1}), namely that $I_{M}(\rho _{1}^{(N)}(t),%
\mathbf{b})$ cannot have negative values for arbitrary $t\geq t_{o}$. To
reach such a result the following proposition is established.

\textbf{THM. 1 - Non-negativity of }$I_{M}(\rho _{1}^{(N)}(t),\mathbf{b})$

\emph{Let us assume that }$\rho _{1}^{(N)}(\mathbf{x}_{1},t)$\emph{\ is an
arbitrary stochastic particular solution of the Master kinetic equation (\ref%
{App-1}) such that the integral }$K_{M}(\rho _{1o}^{(N)}(\mathbf{x}_{1}),%
\mathbf{b})$ \emph{is non-vanishing. Then it follows necessarily that:}

\begin{itemize}
\item \emph{Proposition P1}$_{1}:$%
\begin{equation}
K_{M}(\rho _{1o}^{(N)}(\mathbf{x}_{1}),\mathbf{b})>0.  \label{P3-1-1}
\end{equation}

\item \emph{Proposition P1}$_{2}:$ \emph{for all }$t\in I$\emph{\ with }$%
t>t_{o}$%
\begin{equation}
K_{M}(\rho _{1}^{(N)}(t),\mathbf{b})\geq 0.  \label{P3-1-2}
\end{equation}

\item \emph{Proposition P1}$_{3}:$ \emph{then necessarily the inequality}%
\begin{equation}
I_{M}(\rho _{1}^{(N)}(t),\mathbf{b})\geq 0  \label{P3-1-3}
\end{equation}%
\emph{must hold too.}

\item \emph{Proposition P1}$_{4}:$ \emph{Finally the following necessary and
sufficient condition holds at a given time }$t\in I$ \emph{with} $t\geq
t_{o}:$%
\begin{equation}
K_{M}(\rho _{1}^{(N)}(t),\mathbf{b})=0\Leftrightarrow \rho _{1}^{(N)}(%
\mathbf{x}_{1},t)\equiv \rho _{1M}^{(N)}(\mathbf{v}_{1}).  \label{P3-1-4}
\end{equation}
\end{itemize}

\emph{Proof - }One first notices that $K_{M}(\rho _{1}^{(N)}(t),\mathbf{b})$
can be equivalently written in the form%
\begin{equation}
K_{M}(\rho _{1}^{(N)}(t),\mathbf{b})\equiv -\int\limits_{\Gamma _{1(1)}}d%
\mathbf{x}_{1}\overline{\Theta }_{1}^{(\partial \Omega )}(\overline{\mathbf{r%
}})M(\mathbf{v}_{1},\mathbf{b})k_{1}^{(N)}(\mathbf{r}_{1},t)\frac{\partial
^{2}\widehat{\rho }_{1}^{(N)}(\mathbf{x}_{1},t)}{\partial \mathbf{r}%
_{1}\cdot \partial \mathbf{r}_{1}}.
\end{equation}%
Hence integrating by parts, noting that the gradient term $\frac{\partial
\overline{\Theta }_{1}^{(\partial \Omega )}(\overline{\mathbf{r}})}{\partial
\mathbf{r}_{1}}$ gives a vanishing contribution to the phase-space integral
and upon invoking Eq.(\ref{App-X1}) reported in Appendix B it follows%
\begin{eqnarray}
&&\left. K_{M}(\rho _{1}^{(N)}(t),\mathbf{b})\equiv \int\limits_{\Gamma
_{1(1)}}d\mathbf{x}_{1}\overline{\Theta }_{1}^{(\partial \Omega )}(\overline{%
\mathbf{r}})M(\mathbf{v}_{1},\mathbf{b})\frac{\partial k_{1}^{(N)}(\mathbf{r}%
_{1},t)}{\partial \mathbf{r}_{1}}\cdot \frac{\partial \widehat{\rho }%
_{1}^{(N)}(\mathbf{x}_{1},t)}{\partial \mathbf{r}_{1}}=\right.  \notag \\
&=&\left( N-1\right) \int\limits_{\Gamma _{1(1)}}d\mathbf{x}_{1}\overline{%
\Theta }_{1}^{(\partial \Omega )}(\overline{\mathbf{r}})M(\mathbf{v}_{1},%
\mathbf{b})\frac{\partial \widehat{\rho }_{1}^{(N)}(\mathbf{x}_{1},t)}{%
\partial \mathbf{r}_{1}}\cdot \int\limits_{\Gamma _{1(2)}}d\mathbf{x}_{2}%
\mathbf{n}_{12}\delta \left( \left\vert \mathbf{r}_{2}-\mathbf{r}%
_{1}\right\vert -\sigma \right) \times  \notag \\
&&\widehat{\rho }_{1}^{(N)}(\mathbf{x}_{2},t)k_{2}^{(N)}(\mathbf{r}_{1},%
\mathbf{r}_{2},t).
\end{eqnarray}

Now noting that $\mathbf{n}_{12}\delta \left( \left\vert \mathbf{r}_{2}-%
\mathbf{r}_{1}\right\vert -\sigma \right) =-\frac{\partial }{\partial
\mathbf{r}_{2}}\overline{\Theta }\left( \left\vert \mathbf{r}_{2}-\mathbf{r}%
_{1}\right\vert -\sigma \right) $ and ignoring again a vanishing
contribution carried by $\frac{\partial }{\partial \mathbf{r}_{2}}\overline{%
\Theta }_{2}^{(\partial \Omega )}(\overline{\mathbf{r}})$, the rhs of
previous equation can once more be integrated by parts yielding%
\begin{equation}
K_{M}(\rho _{1}^{(N)}(t),\mathbf{b})\equiv K_{M}^{(1)}(\rho _{1}^{(N)}(%
\mathbf{x}_{1},t),\mathbf{b})+\Delta K_{M}^{(1)}(\rho _{1}^{(N)}(\mathbf{x}%
_{1},t),\mathbf{b}).
\end{equation}%
Hence in terms of the total directional kinetic energy carried by particles $%
1$ and $2,$ namely $M(\mathbf{v}_{1},\mathbf{v}_{2},\mathbf{b})$ (see Eq.(%
\ref{TOTAL DIRECTIONAL ENERGY})) the functional $K_{M}^{(1)}(\rho
_{1}^{(N)}(t),\mathbf{b})$ takes the form%
\begin{equation}
\begin{array}{c}
K_{M}^{(1)}(\rho _{1}^{(N)}(\mathbf{x}_{1},t),\mathbf{b})=\left( N-1\right)
\int\limits_{\Gamma _{1(1)}}d\mathbf{x}_{1}\int\limits_{\Gamma _{1(2)}}d%
\mathbf{x}_{2}\overline{\Theta }_{1}^{(\partial \Omega )}(\overline{\mathbf{r%
}})\overline{\Theta }_{2}^{(\partial \Omega )}(\overline{\mathbf{r}})\times
\\
\frac{\partial \widehat{\rho }_{1}^{(N)}(\mathbf{x}_{1},t)}{\partial \mathbf{%
r}_{1}}\cdot \frac{\partial }{\partial \mathbf{r}_{2}}\widehat{\rho }%
_{1}^{(N)}(\mathbf{x}_{2},t)k_{2}^{(N)}(\mathbf{r}_{1},\mathbf{r}_{2},t)M(%
\mathbf{v}_{1},\mathbf{v}_{2},\mathbf{b})\overline{\Theta }\left( \left\vert
\mathbf{r}_{2}-\mathbf{r}_{1}\right\vert -\sigma \right) .%
\end{array}
\label{K^1}
\end{equation}%
Instead the second term $\Delta K_{M}^{(1)}(\rho _{1}^{(N)}(\mathbf{x}%
_{1},t),\mathbf{b})$ reads
\begin{eqnarray}
&&\Delta K_{M}^{(1)}(\rho _{1}^{(N)}(\mathbf{x}_{1},t),\mathbf{b})\equiv
\left( N-1\right) \int\limits_{\Gamma _{1(1)}}d\mathbf{x}_{1}\overline{%
\Theta }_{1}^{(\partial \Omega )}(\overline{\mathbf{r}})\frac{\partial
\widehat{\rho }_{1}^{(N)}(\mathbf{x}_{1},t)}{\partial \mathbf{r}_{1}}\cdot
\int\limits_{\Gamma _{1(2)}}d\mathbf{x}_{2}\times  \notag \\
&&\overline{\Theta }_{2}^{(\partial \Omega )}(\overline{\mathbf{r}})M(%
\mathbf{v}_{1},\mathbf{b})\overline{\Theta }\left( \left\vert \mathbf{r}_{2}-%
\mathbf{r}_{1}\right\vert -\sigma \right) \widehat{\rho }_{1}^{(N)}(\mathbf{x%
}_{2},t)\frac{\partial }{\partial \mathbf{r}_{2}}k_{2}^{(N)}(\mathbf{r}_{1},%
\mathbf{r}_{2},t),  \label{AK^1}
\end{eqnarray}%
with the gradient $\frac{\partial }{\partial \mathbf{r}_{2}}k_{2}^{(N)}(%
\mathbf{r}_{1},\mathbf{r}_{2},t)$ being given by the differential identity (%
\ref{App-X2}) reported in Appendix B. The procedure is analogous to the one
followed above for the calculation of $\frac{\partial k_{1}^{(N)}(\mathbf{r}%
_{1},t)}{\partial \mathbf{r}_{1}}$ and can be iterated at arbitrary order $%
s=1,N-1$ (see Eq.(\ref{App-X3}) in Appendix B). As a result it follows that $%
K_{M}(\rho _{1}^{(N)}(t),\mathbf{b})$ can be represented in terms of a sum
of the form $K_{M}(\rho _{1}^{(N)}(t),\mathbf{b})\equiv
\sum\limits_{j=1,N-1}K_{M}^{(j)}(\rho _{1}^{(N)}(\mathbf{x}_{1},t),\mathbf{b}%
)$ in which each functional $K_{M}^{(j)}(\rho _{1}^{(N)}(\mathbf{x}_{1},t),%
\mathbf{b})$ is non-negative and symmetric. Therefore $K_{M}(\rho
_{1}^{(N)}(t),\mathbf{b})$ takes the form of a non-negative and symmetric
functional of the type
\begin{eqnarray}
&&K_{M}(\rho _{1}^{(N)}(t),\mathbf{b})=\int\limits_{\Gamma _{1(1)}}d\mathbf{x%
}_{1}\int\limits_{\Gamma _{1(2)}}d\mathbf{x}_{2}\overline{\Theta }%
_{1}^{(\partial \Omega )}(\overline{\mathbf{r}})\overline{\Theta }%
_{2}^{(\partial \Omega )}(\overline{\mathbf{r}})M(\mathbf{v}_{1},\mathbf{v}%
_{2},\mathbf{b})\times  \notag \\
&&F(\mathbf{r}_{1},\mathbf{r}_{2},t)\frac{\partial \widehat{\rho }_{1}^{(N)}(%
\mathbf{x}_{1},t)}{\partial \mathbf{r}_{1}}\cdot \frac{\partial }{\partial
\mathbf{r}_{2}}\widehat{\rho }_{1}^{(N)}(\mathbf{x}_{2},t)\overline{\Theta }%
\left( \left\vert \mathbf{r}_{2}-\mathbf{r}_{1}\right\vert -\sigma \right) ,
\label{REPRESENTATION OF KM}
\end{eqnarray}%
with $M(\mathbf{v}_{1},\mathbf{v}_{2},\mathbf{b})$ being the total
directional kinetic energy (\ref{TOTAL DIRECTIONAL ENERGY}) and $F(\mathbf{r}%
_{1},\mathbf{r}_{2},t)$ a suitable real scalar kernel which is symmetric in
the variables $\mathbf{r}_{1}$ and $\mathbf{r}_{2}.$ This proves validity of
the inequality (\ref{P3-1-3}) (Proposition P1$_{1}$). As a result, invoking
Eq.(\ref{MKI-6}) it follows that the inequalities (\ref{P3-1-1}) and (\ref%
{P3-1-2}), and hence Propositions P1$_{2}$ and P1$_{3}$ manifestly hold too.
Finally, one notices that $K_{M}(\rho _{1}^{(N)}(t),\mathbf{b})=0$ if and
only if identically $\frac{\partial }{\partial \mathbf{r}_{1}}\widehat{\rho }%
_{1}^{(N)}(\mathbf{x}_{1},t)\equiv 0$. Since $\widehat{\rho }_{1}^{(N)}(%
\mathbf{x}_{1},t)$ is by construction a solution of the Master kinetic
equation this requires necessarily that (\ref{P3-1-4}) must hold too
(Proposition P1$_{4})$. \textbf{Q.E.D.}

\bigskip

\subsection{2B - Proof of PMI for the Master kinetic equation}

The next step is to prove the monotonic time-decreasing behavior of the MKI
functional, which involves \emph{No.\#3 }and \emph{No.\#4 MKI Prescriptions }%
and consequently also the validity of \emph{No.\#2 MKI Prescription.} The
first two refer respectively to the validity of the time derivative
inequality (\ref{MKI-2}) and the conditions of existence of kinetic
equilibrium (\ref{MKI-4a}), while the latter one concerns the right-hand
inequality $I_{M}(\rho _{1}^{(N)}(t))\leq 1$. In order to prove these
properties let us preliminarily determine the variation of the total
directional kinetic energy $M(\mathbf{v}_{1},\mathbf{v}_{2},\mathbf{b})$
(see Eq.(\ref{TOTAL DIRECTIONAL ENERGY})) across a binary collision, namely
the quantity $\Delta M(\mathbf{v}_{1},\mathbf{v}_{2},\mathbf{b})\equiv M(%
\mathbf{v}_{1}^{(+)},\mathbf{v}_{2}^{(+)},\mathbf{b})-M(\mathbf{v}_{1}%
\mathbf{v}_{2},\mathbf{b})$. One obtains%
\begin{equation}
\Delta M(\mathbf{v}_{1},\mathbf{v}_{2},\mathbf{b})=\mathbf{b\cdot n}%
_{12}\left\vert \mathbf{n}_{12}\cdot \mathbf{v}_{12}^{(+)}\right\vert
\mathbf{v}_{12}^{(+)}\cdot \mathbf{b-}\left( \mathbf{b\cdot n}_{12}\right)
^{2}\left( \mathbf{n}_{12}\cdot \mathbf{v}_{12}^{(+)}\right) ^{2},
\label{directional eneergy VARIATION}
\end{equation}%
the rhs being expressed in terms of the outgoing particle velocities $(%
\mathbf{v}_{1}^{(+)},\mathbf{v}_{2}^{(+)})$. Then the following proposition
holds.

\bigskip

\textbf{THM. 2 - Property of macroscopic irreversibility }(\textbf{Master
equation PMI theorem})

\emph{Let us assume that }$\rho _{1}^{(N)}(\mathbf{r}_{1},\mathbf{v}_{1},t)$%
\emph{\ is an arbitrary stochastic particular solution of the Master kinetic
equation (\ref{App-1}) such that the integral }$K_{M}(\rho _{1o}^{(N)}(%
\mathbf{x}_{1}),\mathbf{b})$ \emph{is non-vanishing. Then it follows that}

\begin{itemize}
\item \emph{Proposition P2}$_{1}$: \emph{one finds that for all }$t\geq
t_{o}:$%
\begin{eqnarray}
&&\left. \frac{\partial }{\partial t}K_{M}(\rho _{1}^{(N)}(t),\mathbf{b}%
)=-(N-1)\sigma ^{2}\int\limits_{U_{1(1)}}d\mathbf{v}_{1}\int%
\limits_{U_{1(2)}}d\mathbf{v}_{2}\int\limits_{\Omega }d\mathbf{r}%
_{1}\int^{(-)}d\mathbf{\Sigma }_{21}\left\vert \mathbf{v}_{12}^{(+)}\cdot
\mathbf{n}_{12}\right\vert \right.  \notag \\
&&\left( \mathbf{b\cdot n}_{12}\right) ^{2}\left( \mathbf{n}_{12}\cdot
\mathbf{v}_{12}^{(+)}\right) ^{2}\frac{\partial \widehat{\rho }_{1}^{(N)}(%
\mathbf{r}_{1},\mathbf{v}_{1}^{(+)},t)}{\partial \mathbf{r}_{1}}\cdot \frac{%
\partial }{\partial \mathbf{r}_{2}}\widehat{\rho }_{1}^{(N)}(\mathbf{r}_{2}=%
\mathbf{r}_{1}+\sigma \mathbf{n}_{21},\mathbf{v}_{2}^{(+)}t)  \notag \\
&&\left. k_{2}^{(N)}(\mathbf{r}_{1},\mathbf{r}_{2}=\mathbf{r}_{1}+\sigma
\mathbf{n}_{21},t)\leq 0.\right.  \label{P2-2-2}
\end{eqnarray}

\item \emph{Proposition P2}$_{2}$: \emph{the inequality }%
\begin{equation}
\frac{\partial }{\partial t}I_{M}(\rho _{1}^{(N)}(t),\mathbf{b})\leq 0
\label{P2-2-1}
\end{equation}%
\emph{holds identically for all }$t\geq t_{o}$ \emph{so that necessarily}
\begin{equation}
I_{M}(\rho _{1}^{(N)}(t),\mathbf{b})\leq 1.  \label{P2-2-1bis}
\end{equation}

\item \emph{Proposition P2}$_{3}$: \emph{one finds that a given time }$t\in
I $ \emph{with} $t\geq t_{o}:$%
\begin{equation*}
\frac{\partial }{\partial t}K_{M}(\rho _{1}^{(N)}(t),\mathbf{b}%
)=0\Leftrightarrow \rho _{1}^{(N)}(\mathbf{x}_{1},t)\equiv \rho _{1M}^{(N)}(%
\mathbf{v}_{1}).
\end{equation*}%
\emph{Proof - }Upon time-differentiation of the functional $K_{M}(\rho
_{1}^{(N)}(t),\mathbf{b})$ and invoking the first form of the Master kinetic
equation (see Eq.(\ref{App-0}) in Appendix A) one obtains
\begin{eqnarray}
\frac{\partial }{\partial t}K_{M}(\rho _{1}^{(N)}(t),\mathbf{b})
&=&-\int\limits_{\Gamma _{1(1)}}d\mathbf{x}_{1}M(\mathbf{v}_{1},\mathbf{b}%
)k_{1}^{(N)}(\mathbf{r}_{1},t)\left( -\mathbf{v}_{1}\cdot \frac{\partial }{%
\partial \mathbf{r}_{1}}\right) \frac{\partial ^{2}\widehat{\rho }_{1}^{(N)}(%
\mathbf{x}_{1},t)}{\partial \mathbf{r}_{1}\cdot \partial \mathbf{r}_{1}}
\notag \\
&&-\int\limits_{\Gamma _{1(1)}}d\mathbf{x}_{1}M(\mathbf{v}_{1},\mathbf{b})%
\frac{\partial ^{2}\widehat{\rho }_{1}^{(N)}(\mathbf{x}_{1},t)}{\partial
\mathbf{r}_{1}\cdot \partial \mathbf{r}_{1}}\left( \frac{\partial }{\partial
t}\right) k_{1}^{(N)}(\mathbf{r}_{1},t),
\end{eqnarray}%
namely%
\begin{equation}
\frac{\partial }{\partial t}K_{M}(\rho _{1}^{(N)}(t),\mathbf{b}%
)=-\int\limits_{\Gamma _{1(1)}}d\mathbf{x}_{1}M(\mathbf{v}_{1},\mathbf{b})%
\frac{\partial ^{2}\widehat{\rho }_{1}^{(N)}(\mathbf{x}_{1},t)}{\partial
\mathbf{r}_{1}\cdot \partial \mathbf{r}_{1}}\left( \frac{\partial }{\partial
t}+\mathbf{v}_{1}\cdot \frac{\partial }{\partial \mathbf{r}_{1}}\right)
k_{1}^{(N)}(\mathbf{r}_{1},t).
\end{equation}%
Hence, thanks to the differential identity (\ref{DIFF-identity}) it follows:%
\begin{eqnarray}
&&\left. \frac{\partial }{\partial t}K_{M}(\rho _{1}^{(N)}(t),\mathbf{b}%
)=-(N-1)\int\limits_{\Gamma _{1(1)}}d\mathbf{x}_{1}M(\mathbf{v}_{1},\mathbf{b%
})\int\limits_{\overline{\Gamma }_{1(2)}}d\mathbf{x}_{2}\mathbf{v}_{12}\cdot
\mathbf{n}_{12}\times \right.  \notag \\
&&\delta (\left\vert \mathbf{r}_{1}-\mathbf{r}_{2}\right\vert -\sigma
)k_{2}^{(N)}(\mathbf{r}_{1},\mathbf{r}_{2},t)\frac{\partial ^{2}\widehat{%
\rho }_{1}^{(N)}(\mathbf{x}_{1},t)}{\partial \mathbf{r}_{1}\cdot \partial
\mathbf{r}_{1}}\widehat{\rho }_{1}^{(N)}(\mathbf{x}_{2},t).
\end{eqnarray}%
Performing an integration by parts and upon invoking the first differential
identity (\ref{App-10a}) (reported in Appendix B) this delivers:%
\begin{equation}
\frac{\partial }{\partial t}K_{M}(\rho _{1}^{(N)}(t),\mathbf{b})=W_{M}(\rho
_{1}^{(N)}(t),\mathbf{b}),  \label{P2-2-2a}
\end{equation}%
\begin{eqnarray}
&&\left. W_{M}(\rho _{1}^{(N)}(t),\mathbf{b})\equiv \int\limits_{\Gamma
_{1(1)}}d\mathbf{x}_{1}M(\mathbf{v}_{1},\mathbf{b})\frac{\partial \widehat{%
\rho }_{1}^{(N)}(\mathbf{x}_{1},t)}{\partial \mathbf{r}_{1}}(N-1)\right.
\notag \\
&&\times \int\limits_{\overline{\Gamma }_{1(2)}}d\mathbf{x}_{2}\mathbf{v}%
_{12}\cdot \mathbf{n}_{12}\frac{\partial }{\partial \mathbf{r}_{1}}\left[
\delta (\left\vert \mathbf{r}_{1}\mathbf{-r}_{2}\right\vert -\sigma )\right]
k_{2}^{(N)}(\mathbf{r}_{1},\mathbf{r}_{2}t)\widehat{\rho }_{1}^{(N)}(\mathbf{%
x}_{2},t),  \label{P2-2-2b}
\end{eqnarray}%
where $\frac{\partial }{\partial \mathbf{r}_{1}}\left[ \delta (\left\vert
\mathbf{r}_{1}-\mathbf{r}_{2}\right\vert -\sigma )\right] =-\frac{\partial }{%
\partial \mathbf{r}_{2}}\left[ \delta (\left\vert \mathbf{r}_{1}-\mathbf{r}%
_{2}\right\vert -\sigma )\right] $. Hence performing a further integration
by parts and using the second differential identity (\ref{App-10a})
(Appendix B) the previous equation yields%
\begin{eqnarray}
&&\left. W_{M}(\rho _{1}^{(N)}(t),\mathbf{b})=(N-1)\int\limits_{\Gamma
_{1(1)}}d\mathbf{x}_{1}\int\limits_{\overline{\Gamma }_{1(2)}}d\mathbf{x}_{2}%
\mathbf{v}_{12}\cdot \mathbf{n}_{12}M(\mathbf{v}_{1},\mathbf{v}_{2},\mathbf{b%
})\times \right.  \notag \\
&&\delta (\left\vert \mathbf{r}_{1}-\mathbf{r}_{2}\right\vert -\sigma
)k_{2}^{(N)}(\mathbf{r}_{1},\mathbf{r}_{2},t)\frac{\partial \widehat{\rho }%
_{1}^{(N)}(\mathbf{x}_{1},t)}{\partial \mathbf{r}_{1}}\cdot \frac{\partial }{%
\partial \mathbf{r}_{2}}\widehat{\rho }_{1}^{(N)}(\mathbf{x}_{2},t),
\label{P2-CASO-gen}
\end{eqnarray}

where the symmetry property with respect to the exchange of states $\left(
\mathbf{x}_{1},\mathbf{x}_{2}\right) $ has been invoked. In the previous
equation the integration on the Dirac delta can be performed at once letting
\begin{eqnarray}
&&\left. \int\limits_{\Gamma _{1(1)}}d\mathbf{x}_{1}\int\limits_{\overline{%
\Gamma }_{1(2)}}d\mathbf{x}_{2}\delta (\left\vert \mathbf{r}_{1}-\mathbf{r}%
_{2}\right\vert -\sigma )=\sigma ^{2}\int\limits_{U_{1(1)}}d\mathbf{v}%
_{1}\int\limits_{U_{1(2)}}d\mathbf{v}_{2}\int\limits_{\Omega }d\mathbf{r}%
_{1}\right.  \notag \\
&&\left[ \int^{(+)}d\mathbf{\Sigma }_{21}\left\vert \mathbf{v}_{12}\cdot
\mathbf{n}_{12}\right\vert -\int^{(-)}d\mathbf{\Sigma }_{21}\left\vert
\mathbf{v}_{12}\cdot \mathbf{n}_{12}\right\vert \right] ,
\end{eqnarray}%
where the solid-angle integrations in the two integrals on the rhs are
performed respectively on the outgoing $(+)$ and incoming $(-)$ particles.
Furthermore, it is obvious that thanks to the causal form of MCBC (see Eq.(%
\ref{bbb3}) in Appendix C) the integral $\int^{(+)}d\mathbf{\Sigma }_{21}$
can be transformed to a corresponding integration on $\int^{(-)}d\mathbf{%
\Sigma }_{21}.$ Thus the contributions in the two phase-space integrals only
differ because of the variation $\Delta M(\mathbf{v}_{1},\mathbf{v}_{2},%
\mathbf{b})$ of the total directional kinetic energy of particles $1$ and $%
2. $ This implies that
\begin{eqnarray}
&&\left. W_{M}(\rho _{1}^{(N)}(t),\mathbf{b})=(N-1)\sigma
^{2}\int\limits_{U_{1(1)}}d\mathbf{v}_{1}\int\limits_{U_{1(2)}}d\mathbf{v}%
_{2}\int\limits_{\Omega }d\mathbf{r}_{1}\int^{(-)}d\mathbf{\Sigma }%
_{21}\times \right.  \notag \\
&&\left\vert \mathbf{v}_{12}\cdot \mathbf{n}_{12}\right\vert \Delta M(%
\mathbf{v}_{1},\mathbf{v}_{2},\mathbf{b})\frac{\partial \widehat{\rho }%
_{1}^{(N)}(\mathbf{r}_{1},\mathbf{v}_{1}^{(+)},t)}{\partial \mathbf{r}_{1}}%
\cdot \frac{\partial \widehat{\rho }_{1}^{(N)}(\mathbf{r}_{2}=\mathbf{r}%
_{1}+\sigma \mathbf{n}_{21},\mathbf{v}_{2}^{(+)}t)}{\partial \mathbf{r}_{2}}%
\times  \notag \\
&&\left. k_{2}^{(N)}(\mathbf{r}_{1},\mathbf{r}_{2}=\mathbf{r}_{1}+\sigma
\mathbf{n}_{21},t),\right.  \label{P2-2-2c}
\end{eqnarray}%
where the solid-angle integration is performed on the incoming particles
whereas $\Delta M(\mathbf{v}_{1},\mathbf{v}_{2},\mathbf{b})$ is evaluated in
terms of the outgoing particles $(+)$ and therefore must be identified with
the second equation on the rhs of Eq.(\ref{directional eneergy VARIATION}).
Consider now the dependences in terms of the outgoing particle velocities $%
\mathbf{v}_{1}^{(+)}$ and $\mathbf{v}_{2}^{(+)}$ in the previous phase-space
integral. The velocity dependences contained in the factors $\left\vert
\mathbf{v}_{12}\cdot \mathbf{n}_{12}\right\vert $ and $\frac{\partial
\widehat{\rho }_{1}^{(N)}(\mathbf{r}_{1},\mathbf{v}_{1}^{(+)},t)}{\partial
\mathbf{r}_{1}}\cdot \frac{\partial \widehat{\rho }_{1}^{(N)}(\mathbf{r}_{2},%
\mathbf{v}_{2}^{(+)}t)}{\partial \mathbf{r}_{2}}$ are manifestly symmetric
with respect to the variables $\mathbf{v}_{1}^{(+)}$ and $\mathbf{v}%
_{2}^{(+)}.$ On the other hand, as a whole, the same integral should remain
unaffected with respect to the exchange of the outgoing particle velocities $%
\mathbf{v}_{1}^{(+)}\Leftrightarrow \mathbf{v}_{2}^{(+)}.$ This means that
the only term in $\Delta M(\mathbf{v}_{1},\mathbf{v}_{2},\mathbf{b})$ which
can give a non-vanishing contribution is $-\left( \mathbf{b}\cdot \mathbf{n}%
_{12}\right) ^{2}\left( \mathbf{n}_{12}\cdot \mathbf{v}_{12}^{(+)}\right)
^{2}.$ As a consequence the previous integral reduces to%
\begin{eqnarray}
&&\left. \frac{\partial }{\partial t}K_{M}(\rho _{1}^{(N)}(t),\mathbf{b}%
)\equiv W_{M}(\rho _{1}^{(N)}(t),\mathbf{b})=-(N-1)\sigma
^{2}\int\limits_{U_{1(1)}}d\mathbf{v}_{1}\int\limits_{U_{1(2)}}d\mathbf{v}%
_{2}\times \right.  \notag \\
&&\int\limits_{\Omega }d\mathbf{r}_{1}\int^{(-)}d\mathbf{\Sigma }_{21}\frac{%
\partial \widehat{\rho }_{1}^{(N)}(\mathbf{r}_{1},\mathbf{v}_{1}^{(+)},t)}{%
\partial \mathbf{r}_{1}}\cdot \frac{\partial \widehat{\rho }_{1}^{(N)}(%
\mathbf{r}_{2}=\mathbf{r}_{1}+\sigma \mathbf{n}_{21},\mathbf{v}_{2}^{(+)}t)}{%
\partial \mathbf{r}_{2}}\times  \notag \\
&&\left. \left\vert \mathbf{v}_{12}^{(+)}\cdot \mathbf{n}_{12}\right\vert
\left( \mathbf{b\cdot n}_{12}\right) ^{2}\left( \mathbf{n}_{12}\cdot \mathbf{%
v}_{12}^{(+)}\right) ^{2}k_{2}^{(N)}(\mathbf{r}_{1},\mathbf{r}_{2}=\mathbf{r}%
_{1}+\sigma \mathbf{n}_{21},t)\leq 0,\right.  \label{P2-2-2d}
\end{eqnarray}%
(\ref{P2-2-2d})and hence is necessarily negative or null, the second case
occurring only if $\frac{\partial \widehat{\rho }_{1}^{(N)}(\mathbf{r}_{1},%
\mathbf{v}_{1}^{(+)},t)}{\partial \mathbf{r}_{1}}\equiv 0.$ The proof of
Proposition P2$_{2}$ is straightforward since $\frac{\partial }{\partial t}%
I_{M}(\rho _{1}^{(N)}(t),\mathbf{b})\equiv \frac{1}{K_{Mo}}\frac{\partial }{%
\partial t}K_{M}(\rho _{1}^{(N)}(t),\mathbf{b})$ so that the inequality (\ref%
{P2-2-2d}) implies Eq.(\ref{P2-2-1}) and (\ref{P2-2-1bis}) too.\textbf{\ }%
Finally, since $\widehat{\rho }_{1}^{(N)}(\mathbf{r}_{1},\mathbf{v}_{1},t)$
is a solution of the Master kinetic equation $\frac{\partial }{\partial
\mathbf{r}_{1}}\widehat{\rho }_{1}^{(N)}(\mathbf{x}_{1},t)\equiv 0$ occurs
if and only if $\widehat{\rho }_{1}^{(N)}(\mathbf{x}_{1},t)$ coincides with
a Maxwellian kinetic equilibrium of the type (\ref{MAXWELLIAN-PDF}). This
proves\ also Proposition P2$_{3}.$ \textbf{Q.E.D.}
\end{itemize}

\subsection{2C - Proof of the DKE property for the Master kinetic equation}

Let us now show that in validity of THMs. 1 and 2 the time-evolved $\rho
_{1}^{(N)}(\mathbf{x}_{1},t)$ necessarily must decay asymptotically for $%
t-t_{o}\rightarrow +\infty $ to kinetic equilibrium, i.e., that the limit
function $\lim_{t-t_{o}\rightarrow +\infty }\rho _{1}^{(N)}(\mathbf{x}%
_{1},t)\equiv \rho _{1\infty }^{(N)}(\mathbf{x}_{1})$ exists and it
necessarily coincides with a Maxwellian kinetic equilibrium of the type (\ref%
{MAXWELLIAN-PDF}). In this regard the following proposition holds.

\textbf{THM. 3 - Asymptotic behavior of }$I_{M}(\rho _{1}^{(N)}(t),\mathbf{b}%
)$ (\textbf{Master equation-DKE theorem})

\emph{Let us assume that the initial condition} $\rho _{1o}^{(N)}(\mathbf{x}%
_{1})\in \left\{ \rho _{1o}^{(N)}(\mathbf{x}_{1})\right\} $ is\emph{\ such
that the corresponding functional} $K_{M}(\rho _{1o}^{(N)}(\mathbf{x}_{1}),%
\mathbf{b})$ \emph{is non-vanishing, i.e., in view of THM.1 necessarily }$>0$%
\emph{. Then is follows that the corresponding time-evolved solution of the
Master kinetic equation} $\rho _{1}^{(N)}(\mathbf{x}_{1},t)$ \emph{in the
limit }$t-t_{o}\rightarrow +\infty $ \emph{necessarily must decay to kinetic
equilibrium, i.e.,}
\begin{equation}
\lim_{t-t_{o}\rightarrow +\infty }\rho _{1}^{(N)}(\mathbf{x}_{1},t)=\rho
_{1M}^{(N)}(\mathbf{v}_{1}).  \label{P3-3-1}
\end{equation}%
\emph{Proof - }In order to reach the thesis it is sufficient to prove that
necessarily%
\begin{equation}
\lim_{t-t_{o}\rightarrow +\infty }\frac{\partial }{\partial t}I_{M}(\rho
_{1}^{(N)}(t),\mathbf{b})=0.  \label{EQQ.1}
\end{equation}%
In fact, let us assume "\textit{ad absurdum}" that $\frac{\partial }{%
\partial t}I_{M}(\rho _{1}^{(N)}(t),\mathbf{b})\leq -k^{2}$ with $k^{2}>0$ a
real constant. Then THM.2 (proposition P2$_{2}$) requires that%
\begin{equation}
\lim_{t-t_{o}\rightarrow +\infty }I_{M}(\rho _{1}^{(N)}(t),\mathbf{b})\leq
-\lim_{t-t_{o}\rightarrow +\infty }(t-t_{o})k^{2}=-\infty ,
\end{equation}%
a result which contradicts THM.1. This proves the validity of Eq.(\ref{EQQ.1}%
). Furthermore, by construction $\frac{\partial }{\partial t}I_{M}\equiv
\frac{1}{K_{Mo}}\frac{\partial }{\partial t}K_{M}$ and furthermore $\frac{%
\partial }{\partial t}K_{M}$ is identified with the functional $W_{M}(\rho
_{1}^{(N)}(t),\mathbf{b})\leq 0$ which is determined by Eq.(\ref{P2-2-2d}).
At this point one notices that, thanks to continuity of the functional $%
W_{M}(\rho _{1}^{(N)}(t),\mathbf{b}),$ necessarily the identity
\begin{equation}
\lim_{t-t_{o}\rightarrow +\infty }\frac{\partial }{\partial t}I_{M}(\rho
_{1}^{(N)}(\mathbf{x}_{1},t),\mathbf{b})=W_{M}(\rho _{1\infty }^{(N)}(%
\mathbf{x}_{1}),\mathbf{b})
\end{equation}%
holds where, thanks to global existence of the $1-$body PDF (see Ref.\cite%
{noi7}), the limit function
\begin{equation}
\lim_{t-t_{o}\rightarrow +\infty }\rho _{1}^{(N)}(\mathbf{x}_{1},t)\equiv
\rho _{1\infty }^{(N)}(\mathbf{x}_{1})
\end{equation}%
necessarily exists. As a consequence Eq.(\ref{EQQ.1}) requires also the
equation%
\begin{equation}
W_{M}(\rho _{1\infty }^{(N)}(\mathbf{x}_{1}),\mathbf{b})=0
\end{equation}%
to hold. Upon invoking proposition P2$_{4}$ of THM.2 this implies that
necessarily $\rho _{1\infty }^{(N)}(\mathbf{x}_{1})=\rho _{1M}^{(N)}(\mathbf{%
v}_{1})$ so the thesis\ (\ref{P3-3-1}) is proved. Incidentally, thanks to
THM. 1, this requires also that%
\begin{equation}
\lim_{t-t_{o}\rightarrow +\infty }I_{M}(\rho _{1}^{(N)}(t),\mathbf{b}%
)=I_{M}(\rho _{1\infty }^{(N)}(\mathbf{x}_{1}),\mathbf{b})=0.
\end{equation}

\textbf{Q.E.D}.

\subsection{2D - Remarks}

A few remarks are worth being pointed out regarding the results presented
above.

\begin{enumerate}
\item \emph{Remark \#1:} The choice of the MKI functional considered here
(see Eq.(\ref{MKI-functional-1})) is just one of the infinite particular
admissible realizations which meet the complete set of MKI-prescriptions
indicated above. In particular the choice of the velocity moment $M(\mathbf{v%
}_{1},\mathbf{b})$ considered here (see Eq.(\ref{MOMENT-1})) remains in
principle arbitrary, since $\left\vert \mathbf{v}_{1}\cdot \mathbf{b}%
\right\vert ^{2}$ can be equivalently replaced, for example, by any factor
of the form $\left\vert \mathbf{v}_{1}\cdot \mathbf{b}\right\vert ^{2n},$
with $n\geq 1.$ Furthermore it is obvious that $M(\mathbf{v}_{1},\mathbf{b})$
can be replaced by any function of the form $M(\mathbf{v}_{1},\mathbf{b}%
)+\Delta M(\mathbf{v}_{1},\mathbf{b})$, being $\Delta M(\mathbf{v}_{1},%
\mathbf{b})$ prescribed in such a way that its contribution to $\frac{%
\partial }{\partial t}I_{M}$ vanishes identically so that the validity of
the inequality (\ref{P2-2-1}) in THM. 2 is preserved. This implies in turn
that the prescription of the MKI functional $I_{M}(\rho _{1}^{(N)}(t))$
remains in principle non-unique.

\item \emph{Remark \#2:} A possible issue is related to the requirement that
the renormalized $1-$body PDF, as the $1-$body PDF itself, are strictly
positive at all times and are non-vanishing. Here it is sufficient to state
that an elementary consequence of the theory of the Master kinetic equation
developed in Ref.\cite{noi3} is that, provided the corresponding initial $N-$%
body PDF set at a prescribed initial time $t_{o}$ is strictly positive in
the whole $N-$body phase-space, both the corresponding renormalized $1-$body
PDF, as the $1-$body PDF remain necessarily strictly positive too at all
times and everywhere in the $1-$body phase-space.

\item \emph{Remark \#3:} It must be stressed that the signature of the time
derivative $\frac{\partial }{\partial t}I_{M}(\rho _{1}^{(N)}(t),\mathbf{b})$
actually depends crucially on the adoption of the causal form of MCBC (i.e.,
see Eq.(\ref{bbb1}) or (\ref{bbb3}) in Appendix C) rather than the
anti-causal one (given instead by Eq.(\ref{bbb2})). The first choice of
course is mandatory in view of the causality principle. Indeed, it is
immediate to prove that $\frac{\partial }{\partial t}I_{M}(\rho
_{1}^{(N)}(t),\mathbf{b})$ changes signature if the anti-causal MCBC Eq.(\ref%
{bbb2}) is invoked.

\item \emph{Remark \#4:} THM.2 warrants that Macroscopic irreversibility,
namely the inequality $\frac{\partial }{\partial t}I_{M}(\rho _{1}^{(N)}(t),%
\mathbf{b})\leq 0$ occurs specifically because of: a) the time-variation of
the $\mathbf{b}-$directional total kinetic energy which occurs at arbitrary
binary collision events; b) the occurrence of a velocity-space anisotropy in
the $1-$body PDF, i.e., the fact that the same PDF may not coincide with a
local Maxwellian PDF.

\item \emph{Remark \#5: }The existence of the limit function $%
\lim_{t\rightarrow +\infty }\rho _{1}^{(N)}(\mathbf{x}_{1},t)=\rho _{1\infty
}^{(N)}(\mathbf{x}_{1})$ follows uniquely as a consequence of the global
existence theorem holding for the Master kinetic equation \cite{noi7}.

\item \emph{Remark \#6: }Last but not least, the fact that the same limit
function may coincide or not with the Maxwellian kinetic equilibrium (\ref%
{MAXWELLIAN-PDF}) depends specifically on the functional setting prescribed
for the same PDF $\rho _{1}^{(N)}(\mathbf{x}_{1},t).$ More precisely DKE can
only occur provided $\rho _{1}^{(N)}(\mathbf{x}_{1},t)$ is a suitably-smooth
stochastic PDF such that the MKI functional exists\ for the corresponding
initial PDF at time $t_{o},$ i.e., $\rho _{1}^{(N)}(\mathbf{x}%
_{1},t_{o})=\rho _{1o}^{(N)}(\mathbf{x}_{1}).$
\end{enumerate}

\bigskip

THMs 1-3 represent the main results of the PMI/DKE theory developed here. In
particular, they show that the notion of macroscopic irreversibility and
that of decay to kinetic equilibrium are intimately connected. The crucial
issues which remain to be addressed are whether these phenomena are actually
consistent with the fundamental symmetry properties of the underlying
Boltzmann-Sinai CDS and to analyze the physical origin and implications of
the present theory.

A detailed discussion on these topics is given below in the following two
sections.

\section{3 - Consistency of MPI/DKE theory with microscopic dynamics}

The problem to be posed in the present section concerns the investigation of
consistency between the occurrence of the MPI/DKE phenomenon and the
time-evolution of the underlying time-reversible, conservative and energy
conserving $N-$body Boltzmann-Sinai classical dynamical system $S_{N}-$CDS.

\begin{enumerate}
\item \emph{First issue: consistency with the microscopic reversibility
principle} - This is related to the famous objection raised by Loschmidt to
the Boltzmann equation and Boltzmann H-theorem: i.e., \emph{whether} and
possibly also \emph{how} it may be possible to reconcile the validity of the
reversibility principle for the $S_{N}-$CDS with the manifestation of a
decay of the $1-$body PDF to kinetic equilibrium, i.e., the uniform
Maxwellian PDF of the form (\ref{MAXWELLIAN-PDF}), as predicted by the above
Master equation-DKE Theorem. That a satisfactory answer to this question is
actually possible follows from elementary considerations which are based on
the axiomatic \textquotedblleft \textit{ab initio}\textquotedblright\
statistical description realized by the Master kinetic equation.\ In this
regard it is worth recalling the discussion reported above concerning the
role of MCBC regarding the functional $\frac{\partial }{\partial t}%
I_{M}(\rho _{1}^{(N)}(t)).$ In particular, it is obvious that the signature
depends on whether the causal (or anti-causal) form of MCBC is invoked (see
Appendix C). Such a choice is not arbitrary since, for consistency with the
causality principle, it must depend on the microscopic arrow of time,\ i.e.,
the orientation of the time axis chosen for the reference frame. Based on
these premises, consistency between the occurrence of macroscopic
irreversibility associated with the DKE phenomenon and the principle of
microscopic reversibility can immediately be established. Indeed, it is
sufficient to notice that when a time-reversal or a velocity-reversal is
performed on the $S_{N}$ $-$CDS the form of the collision boundary
conditions (i.e., in the present case the MCBC provided by Eq.(\ref{bbb1})
in Appendix C) must be changed, replacing them with the corresponding
anti-causal ones, i.e., Eq.(\ref{bbb2}). This manifestly implies that MKI
functional decreases in both cases, i.e., after performing the
time-reversal, so that no contradiction can possibly arise in this case
between THM.3 and the microscopic reversibility principle.

\item \emph{Second issue: consistency with Poincare' recurrence theorem
(PRT) - }Similar considerations concern the consistency with PRT as well as
the conservation of total (kinetic) energy for the $S_{N}$ $-$CDS (see also
the related Zermelo's objection in the Introduction). In fact, first, as
shown in Ref.\cite{noi5} by construction the Master collision operator
admits the customary Boltzmann collisional invariants, including total
kinetic energy of colliding particles. Hence, total energy conservation is
again warranted for $S_{N}$ $-$CDS. Second, regarding PRT, it concerns the
Lagrangian phase-space trajectories of the $S_{N}$\ $-$CDS, i.e., the fact
that almost all of these trajectories return arbitrarily close - in a
suitable sense to be prescribed in terms of a distance defined on the $N-$%
body phase-space - to their initial condition after a suitably large
"recurrence time".\textbf{\ }Incidentally, its magnitude depends strongly
both on the same initial condition and the notion of distance to be
established on the same phase-space. Nevertheless, such a "recurrence
effect" influences only the Lagrangian time evolution of the $N-$body PDF
which occurs along the same Lagrangian $N-$body phase-space trajectories.
Instead, the same recurrence effect has manifestly no influence on the time
evolution of the Eulerian $1-$body PDF which is advanced in time in terms of
the Eulerian kinetic equation represented by the Master kinetic equation.
Therefore the mutual consistency of DKE and PRT remains obvious.
\end{enumerate}

Hence, in the framework of the axiomatic \textquotedblleft \textit{ab initio}%
\textquotedblright\ statistical theory based on the Master kinetic equation
the full consistency is warranted with the microscopic dynamics of the
underlying Boltzmann-Sinai CDS.

\bigskip

\section{4 - Physical implications}

Let us now investigate the physical interpretation and main implications
emerging\ from the PMI/DKE theory developed here. The first issue\ is
related to the physical mechanism at the basis of the PMI/DKE phenomenology.

It is well known that in the context of Boltzmann kinetic theory the
property of macroscopic irreversibility as well as the occurrence of the
DKE-phenomenon are both determined by the Boltzmann H-theorem. As recalled
above, this is expressed in terms of the production rate for the
Boltzmann-Shannon entropy $\frac{\partial }{\partial t}S(\rho _{1}(t)),$
with $S(\rho _{1}(t))$ being interpreted as a measure of the ignorance
associated with a solution of the Boltzmann equation.\ In fact the customary
interpretation is that they arise specifically because of the validity of
the entropic inequality (\ref{Entropic enequality}), i.e., the monotonic
increase of $S(\rho _{1}(t))$, and the corresponding entropic equality (\ref%
{entropic equality}) stating a necessary and sufficient condition for
kinetic equilibrium. Such a theorem is actually intimately related with the
equation itself. In fact both the theorem and the equation generally hold
only for stochastic PDFs $\rho _{1}(t)=\rho _{1}(\mathbf{x}_{1},t)$ which
are suitably-smooth and not for distributions \cite{noi1}. According to
Boltzmann's original interpretation, however, both the Boltzmann equation
and Boltzmann H-theorem should only hold when the so-called Boltzmann-Grad
limit is invoked, i.e. based\ on the limit operator $L_{BG}\equiv \lim
_{\substack{ N\rightarrow +\infty  \\ N\sigma ^{2}\sim O(1)}}$ (see Ref.
\cite{noi3,noi7,noi10}).

In striking departure from such a picture:

\begin{itemize}
\item The axiomatic "\textit{ab initio}" theory based on the Master kinetic
equation and the present PMI/DKE theory are applicable to an arbitrary
finite Boltzmann-Sinai CDS. This means that they hold for hard-sphere
systems having a finite number of particles and with finite diameter and
mass, i.e., without the need of invoking validity of asymptotic conditions.

\item The main departure with respect to Boltzmann kinetic theory arises
because, as earlier discovered \cite{noi6},\ the Boltzmann-Shannon entropy
associated with an arbitrary stochastic $1-$body PDF $\rho
_{1}^{(N)}(t)=\rho _{1}^{(N)}(\mathbf{x}_{1},t)$ solution of the Master
kinetic equation is identically conserved. Thus both PMI and DKE are
essentially unrelated to the Boltzmann-Shannon entropy.

\item In the case of the Master kinetic equation the physical mechanism
responsible for the occurrence of both PMI and DKE is unrelated with the
Boltzmann-Shannon entropy. In fact, as shown here, it arises because of the
properties of the MKI functional $I_{M}(\rho _{1}^{(N)}(t),\mathbf{b})$ when
it is expressed in terms of an arbitrary stochastic PDF $\rho
_{1}^{(N)}(t)=\rho _{1}^{(N)}(\mathbf{x}_{1},t)$ solution of the Master
kinetic equation. The only requirement is that the initial PDF $\rho
_{1o}^{\left( N\right) }(\mathbf{x}_{1})$ is prescribed so that the
corresponding MKI functional $I_{M}(\rho _{1o}^{(N)}(\mathbf{x}_{1}),\mathbf{%
b})$ exists.

\item As shown here the MKI functional is a suitably-weighted phase-space
moment of $\rho _{1}^{(N)}(\mathbf{x}_{1},t)$ which can be interpreted as an
information measure for the same PDF, namely belongs to the interval $\left[
0,1\right] ,$ and exhibits a monotonic-decreasing time-dependence, i.e., the
property of macroscopic irreversibility.

\item In addition both $I_{M}(\rho _{1}^{(N)}(t),\mathbf{b})$ and its time
derivative $\frac{\partial }{\partial t}I_{M}(\rho _{1}^{(N)}(t),\mathbf{b})$
vanish identically if and only if the $1-$body PDF coincides with a
Maxwellian kinetic equilibrium of the type (\ref{MAXWELLIAN-PDF}). This
warrants in turn also the occurrence of the DKE-phenomenon for $\rho
_{1}^{(N)}(\mathbf{x}_{1},t)$, i.e., that for $t-t_{o}\rightarrow +\infty $
the same PDF must decay to a Maxwellian kinetic equilibrium of this type.

\item Finally, it is interesting to point out the peculiar behavior of the
MKI functional $I_{M}(\rho _{1}^{(N)}(t),\mathbf{b})$ and its time
derivative $\frac{\partial }{\partial t}I_{M}(\rho _{1}^{(N)}(t),\mathbf{b})$
when the Boltzmann-Grad limit is considered. In particular the $1-$ and $2-$%
body occupation coefficients $k_{1}^{(N)}(\mathbf{r}_{1},t)$ and $%
k_{2}^{(N)}(\mathbf{r}_{1},\mathbf{r}_{2},t)$ which appear in the Master
kinetic equation (see Appendix B, Eqs.(\ref{App-4}) and (\ref{App-5}))
become respectively%
\begin{equation}
\left\{
\begin{array}{c}
L_{BG}k_{1}^{(N)}(\mathbf{r}_{1},t)=1 \\
L_{BG}k_{2}^{(N)}(\mathbf{r}_{1},\mathbf{r}_{2},t)=1%
\end{array}%
.\right.
\end{equation}%
As a consequence the limit functionals $L_{BG}I_{M}(\rho _{1}^{(N)}(t),%
\mathbf{b})$ and $L_{BG}\frac{\partial }{\partial t}I_{M}(\rho _{1}^{(N)}(t),%
\mathbf{b}),$ are necessarily identically vanishing. This means that the
present theory applies properly when the exact Master kinetic equation is
considered and not to its asymptotic approximation obtained in the
Boltzmann-Grad limit, namely the Boltzmann kinetic equation (see Refs.\cite%
{noi3,noi7}).
\end{itemize}

An interesting issue, in the context of the PMI/DKE theory for the Master
kinetic equation, is the role of MCBC in giving rise to the phenomena of
macroscopic irreversibility and decay to kinetic equilibrium. Let us analyze
for this purpose the two cases represented by unary and binary hard-sphere
elastic collisions.

First, let us recall the customary treatment of collision boundary
conditions for unary collision events (also referred to as the so-called
mirror reflection CBC; see for example Cercignani \cite%
{Cercignani1969a,Cercignani1975}). This refers to the occurrence at a
collision time $t_{i}$\ of a single unary elastic collision for particle $1$%
\ at the boundary $\partial \Omega .$\textbf{\ }Let us denote by\textbf{\ }$%
\mathbf{n}_{1}$\textbf{\ }the inward normal to the stationary rigid boundary
$\partial \Omega $\ at the point of contact with the same particle and
respectively $\mathbf{x}_{1}^{(-)}(t_{1})=\left( \mathbf{r}_{1}(t_{1}),%
\mathbf{v}_{1}^{(-)}(t_{1})\right) $ and $\mathbf{x}_{1}^{(+)}(t_{1})=\left(
\mathbf{r}_{1}(t_{1}),\mathbf{v}_{1}^{(+)}(t_{1})\right) $ the incoming and
outgoing particle states while $\mathbf{v}_{1}^{(+)}$ is determined by the
elastic collision law for unary collisions, namely
\begin{equation}
\mathbf{v}_{1}^{(+)}=\mathbf{v}_{1}^{(-)}-2\mathbf{n}_{1}\mathbf{n}_{1}\cdot
\mathbf{v}_{1}^{(-)}.  \label{UNARY - elastic collision law}
\end{equation}%
Then, the PDF-conserving CBC for the $1-$body PDF requires that the
following identity holds%
\begin{equation}
\rho _{1}^{\left( N\right) }(\mathbf{x}_{1}^{(+)}(t_{1}),t_{i})=\rho
^{\left( N\right) }(\mathbf{x}_{1}^{(-)}(t_{i}),t_{i}),
\label{UNARY-PDF-CONSERVING CBC}
\end{equation}%
with $\rho _{1}^{\left( N\right) }(\mathbf{x}_{1}^{(+)}(t_{1}),t_{i})\equiv
\rho _{1}^{\left( N\right) (+)}(\mathbf{x}_{1}^{(+)}(t_{1}),t_{i})$ and $%
\rho ^{\left( N\right) }(\mathbf{x}_{1}^{(-)}(t_{i}),t_{i})\equiv \rho
^{\left( N\right) (-)}(\mathbf{x}_{1}^{(-)}(t_{i}),t_{i})$ denoting the
outgoing and incoming $1-$body PDF respectively. This identifies the
PDF-conserving CBC usually adopted in Boltzmann kinetic theory \cite%
{Boltzmann1972} (Grad \cite{Grad}; see also related discussions in Refs.\cite%
{noi2,noi3,noi4}). The obvious physical implication of Eq.(\ref%
{UNARY-PDF-CONSERVING CBC}) is that $\rho _{1}^{\left( N\right) }(\mathbf{x}%
_{1}^{(+)}(t_{1}),t_{i})$ (and $\rho ^{\left( N\right) }(\mathbf{x}%
_{1}^{(-)}(t_{i}),t_{i}))$ should be necessarily an even function of the
velocity component $\mathbf{n}_{1}\cdot \mathbf{v}_{1}^{(-)}.$ Indeed as
shown in Refs.\cite{noi2,noi3} the PDF-conserving CBC (\ref%
{UNARY-PDF-CONSERVING CBC}) should be replaced with a suitable CBC
identified with the MCBC condition (see also Appendix C). When realized in
terms of its causal form (predicting the outgoing PDF in terms of the
incoming one) the MCBC for unary collisions is just:%
\begin{equation}
\rho _{1}^{\left( N\right) (+)}(\mathbf{x}_{1}^{(+)}(t_{1}),t_{i})=\rho
^{\left( N\right) (-)}(\mathbf{x}_{1}^{(+)}(t_{i}),t_{i}),
\end{equation}%
with $\rho ^{\left( N\right) (-)}(\mathbf{x}_{1}^{(+)}(t_{i}),t_{i})$
denoting the incoming $1-$body PDF evaluated in terms of the outgoing state $%
\mathbf{x}_{1}^{(+)}(t_{i}).$ Assuming left-continuity (see related
discussion in Ref.\cite{noi2}). this can then be identified with $\rho
^{\left( N\right) (-)}(\mathbf{x}_{1}^{(+)}(t_{i}),t_{i})\equiv \rho
^{\left( N\right) }(\mathbf{x}_{1}^{(+)}(t_{i}),t_{i}),$ thus yielding%
\begin{equation}
\rho _{1}^{\left( N\right) (+)}(\mathbf{x}_{1}^{(+)}(t_{1}),t_{i})=\rho
^{\left( N\right) }(\mathbf{x}_{1}^{(+)}(t_{i}),t_{i}).  \label{UNARY MCBC}
\end{equation}%
Eq.(\ref{UNARY MCBC}) provides the physical prescription for the collision
boundary condition, which is referred to as MCBC, holding for the $1-$body
PDF at arbitrary unary collision events. It is immediate to realize that the
function $\rho ^{\left( N\right) }(\mathbf{x}_{1}^{(+)}(t_{i}),t_{i})$ need
not generally be even with respect to the velocity component $\mathbf{n}%
_{1}\cdot \mathbf{v}_{1}^{(-)}.$ In addition Eq.(\ref{UNARY MCBC}), just as (%
\ref{UNARY-PDF-CONSERVING CBC}), also permits the existence of the customary
collisional invariants which in the case of unary collisions are $%
X=1,\left\vert \mathbf{n}_{1}\cdot \mathbf{v}_{1}^{(-)}\right\vert ,\mathbf{v%
}_{1}\cdot \left[ \underline{\underline{\mathbf{1}}}-\mathbf{n}_{1}\mathbf{n}%
_{1}\right] ,$ $v_{1}^{2}.$ As a consequence, one can show that Eq.(\ref%
{UNARY MCBC}) warrants at the same time also the validity of the so-called
no-slip boundary conditions for the fluid velocity field $\mathbf{V}(\mathbf{%
r}_{1},t)$ carried by the $1-$body PDF $\rho _{1}^{\left( N\right) }(\mathbf{%
x}_{1},t)$.

The treatment of MCBC holding for the $2-$body PDF in case of binary
collision events is analogous and is recalled for convenience in Eq.(\ref%
{bbb2}) of Appendix C.

Let us briefly analyze the qualitative physical implications of Eqs.(\ref%
{UNARY MCBC}) and (\ref{bbb2}) as far as the DKE theory is concerned. First,
we notice that unary collisions cannot produce in a proper sense a
velocity-isotropization effect since, as shown by Eq.(\ref{UNARY MCBC}), in
such a case MCBC gives rise only to a change in the velocity distribution
occurring during a unary collision due to a single component of the particle
velocity, namely $\mathbf{n}_{1}\cdot \mathbf{v}_{1}^{(-)}$.\textbf{\ }As a
consequence, this explains why unary collisions do not affect the rate of
change of the MKI functional (see THM.2). Second, Eq.(\ref{bbb2}) shows - on
the contrary - that binary collisions actually do affect by means of MCBC a
velocity-spreading for the $1-$\ and $2-$body PDF.\ In particular, since the
spreading effect occurs in principle for all components of
particle-velocities affecting both particles $1$\ and $2$, this explains why
binary collisions are actually responsible for the irreversible
time-evolution of the MKI functional (see THM.s 2 and 3).

In turn, as implied by THM.3, DKE arises because of the phenomenon of
macroscopic irreversibility (THM.2). The latter arises due specifically to
the possible occurrence of a\textbf{\ }velocity-space anisotropy which
characterizes the $1-$body PDF when the same PDF differs locally from
kinetic equilibrium. In turn, this requires also that the $1-$body PDF
belongs to the functional class of admissible stochastic PDFs\ $\left\{ \rho
_{1}^{(N)}(\mathbf{x}_{1},t)\right\} $. In difference to Boltzmann kinetic
theory, however, the key physical role is actually ascribed to the MKI
functional $I_{M}(\rho _{1}^{(N)}(t))$ rather than the Boltzmann-Shannon
entropy $S_{1}(\rho _{1}^{(N)}(t))$. In fact, as shown in Ref.\cite{noi6}
the same functional remains constant in time once the Master kinetic
equation is adopted. Rather, as shown by THM.2, it is actually the Master
kinetic information $I_{M}(\rho _{1}^{(N)}(t))$\ which exhibits the
characteristic signatures of macroscopic irreversibility.

The key differences arising between the two theories, i.e., the Boltzmann
equation-DKE and the Master equation-DKE, are of course related to the
different and peculiar intrinsic properties of the Boltzmann and Master
kinetic equations. In particular, as discussed at length elsewhere (see Refs.%
\cite{noi1,noi3,noi6,noi7}), precisely because the Boltzmann equation is
only an asymptotic approximation of the Master kinetic equation explains why
a loss of information occurs in Boltzmann kinetic theory and consequently
the related Boltzmann-Shannon entropy is not conserved.

The present investigation shows that in the context of the Master kinetic
equation, the macroscopic irreversibility property, i.e., the monotonic
time-decay behavior of the MKI functional, can be explained at a more
fundamental level, i.e., based specifically on the time-variation of the $%
\mathbf{b}-$directional total kinetic energy which occurs at arbitrary
binary collision events.

The Master equation-DKE theorem\ (THM.3) given above provides a
first-principle proof of the existence of the phenomenon of DKE occurring
for the kinetic description of a finite number of extended hard-spheres,
i.e., described by means of the Master kinetic equation. More precisely, the
DKE phenomenon affects the $1-$body PDFs belonging to the admissible
functional class $\left\{ \rho _{1}^{(N)}(\mathbf{x}_{1},t)\right\} $
determined according to the \emph{MKI Prescription No.\#0}.

\section{5 - Conclusions}

In this paper the problem of the property of microscopic irreversibility
(PMI) and decay to kinetic equilibrium (DKE) have been addressed. In doing
so original ideas and implications are adopted of the new "\textit{ab initio}%
" approach for hard-sphere systems recently developed in the context of
Classical Statistical Mechanics \cite{noi1,noi2}. These are not just small
deviations from standard literature approaches. These developments, in fact,
have opened up a host of exciting new problems and subjects of investigation
in kinetic theory based on the Master kinetic equation for the so-called
Boltzmann-Sinai classical dynamical system (CDS). In fact, the "\textit{ab
initio}" approach, and the present paper in particular, represent an attempt
at providing new foundational bases to the classical statistical mechanics
of hard-sphere systems. The topic which has been pursued here - which
represents also a challenging test for the validity of the new approach -
concerns the investigation of the physical origins of PMI\ and the related
DKE phenomenon arising \emph{in finite }$N-$\emph{body hard-sphere systems.}
These issues refer in particular to:

\begin{itemize}
\item \emph{The proof of the non-negativity of Master kinetic information
(THM.1, subsection 2A) together with the property of macroscopic
irreversibility (PMI; THM.2, subsection 2B).}

\item \emph{The establishment of THM.3 (subsection 2C) and the related proof
of the property of decay to kinetic equilibrium (DKE).}

\item \emph{The consistency of PMI and DKE with microscopic dynamics
(Section 3).}

\item \emph{The analysis of the main physical implications of DKE (Section
4).}
\end{itemize}

The theory presented here departs in several respects from previous
literature and notably from Boltzmann kinetic theory. The main differences
actually arise because of the non-asymptotic character of the new theory,
i.e., the fact that it applies to arbitrary dense or rarefied systems for
which the finite number and size of the constituent particles is accounted
for \cite{noi3}. In this paper basic consequences of the new theory have
been investigated which concern the phenomenon of decay to global kinetic
equilibrium.

The present results are believed to be crucial, besides in mathematical
research, for the physical applications of the "\textit{ab initio}"
statistical theory, i.e., the Master kinetic equation.\ Indeed, regarding
challenging future developments of the theory one should mention among
others the following examples of possible (and mutually-related) routes
worth to be explored. One is related to the investigation of the
time-asymptotic properties of the same kinetic equation, for which the
present paper may represent a useful basis. The second goal refers to the
possible extension of the theory to mixtures formed by hard spheres of
different masses and diameter which possibly undergo both elastic and
anelastic collisions. The third one concerns the investigation of
hydrodynamic regimes for which a key prerequisite is provided by the DKE
theory here established.

\section{Acknowledgments}

Work developed in part within the research projects: A) the Albert Einstein
Center for Gravitation and Astrophysics, Czech Science Foundation No.
14-37086G; B) the research projects of the Czech Science Foundation GA\v{C}R
grant No. 14-07753P; C) the grant No. 02494/2013/RRC \textquotedblleft
\textit{kinetick\'{y} p\v{r}\'{\i}stup k proud\u{e}n\'{\i} tekutin}%
\textquotedblright\ (kinetic approach to fluid flow) in the framework of the
\textquotedblleft Research and Development Support in Moravian-Silesian
Region\textquotedblright , Czech Republic. Initial framework and motivations
of the investigation were based on the research projects developed by the
Consortium for Magnetofluid Dynamics (University of Trieste, Italy) and the
MIUR (Italian Ministry for Universities and Research) PRIN Research Program
\textquotedblleft \textit{Problemi Matematici delle Teorie Cinetiche e
Applicazioni}\textquotedblright , University of Trieste, Italy. One of the
authors (M.T.) is grateful to the International Center for Theoretical
Physics (Miramare, Trieste, Italy) for the hospitality during the
preparation of the manuscript.

\section{Appendix A: Realizations of the Master kinetic equation}

For completeness we recall here the two equivalent forms of the Master
kinetic equation \cite{noi3}. In terms of the renormalized $1-$body PDF $%
\widehat{\rho }_{1}^{(N)}(\mathbf{x}_{1},t)$ (see Eq.(\ref{App-00}) ) the
\emph{first form }of the same equation reads%
\begin{equation}
L_{1\left( 1\right) }\widehat{\rho }_{1}^{(N)}(\mathbf{x}_{1},t)=0,
\label{App-0}
\end{equation}%
with $L_{1\left( 1\right) }=\frac{\partial }{\partial t}+\mathbf{v}_{1}\cdot
\frac{\partial }{\partial \mathbf{r}_{1}}$ denoting the $1-$body
free-streaming operator. Hence it follows
\begin{equation}
L_{1\left( 1\right) }\rho _{1}^{(N)}(\mathbf{x}_{1},t)=\widehat{\rho }%
_{1}^{(N)}(\mathbf{x}_{1},t)L_{1\left( 1\right) }k_{1}^{(N)}(\mathbf{r}%
_{1},t),  \label{APP-0b}
\end{equation}%
where explicit evaluation of the rhs the last equation (see also Eq.(\ref%
{DIFF-identity}) below) yields
\begin{eqnarray}
&&\left. \widehat{\rho }_{1}^{(N)}(\mathbf{x}_{1},t)L_{1\left( 1\right)
}k_{1}^{(N)}(\mathbf{r}_{1},t)=\left( N-1\right) \sigma
^{2}\int\limits_{U_{1(2)}}d\mathbf{v}_{2}\int d\Sigma _{21}\mathbf{v}%
_{21}\cdot \mathbf{n}_{21}\right.  \notag \\
&&\overline{\Theta }^{\ast }(\mathbf{r}_{2})k_{2}^{(N)}(\mathbf{r}_{1},%
\mathbf{r}_{2},t)\widehat{\rho }_{1}^{(N)}(\mathbf{x}_{1},t)\widehat{\rho }%
_{1}^{(N)}(\mathbf{x}_{2},t),  \label{App-01}
\end{eqnarray}%
with $\overline{\Theta }^{\ast }(\mathbf{r}_{2})\equiv \overline{\Theta }%
_{i}^{(\partial \Omega )}(\overline{\mathbf{r}})$ and $k_{2}^{(N)}(\mathbf{r}%
_{1},\mathbf{r}_{2},t)$ being identified with the definitions given
respectively by Eqs. (\ref{App-2b}) and Eq.(\ref{App-4}) in Appendix B. Then
consistent with Ref.\cite{noi3} and upon invoking the causal form of MCBC
(see Eq.(\ref{bbb3}) in Appendix C) the same equation can be equivalent
written in the equivalent \emph{second form} of the Master kinetic equation
\cite{noi3}. The corresponding initial-value problem, taking the form:%
\begin{equation}
\left\{
\begin{array}{c}
L_{1\left( 1\right) }\rho _{1}^{(N)}(\mathbf{x}_{1},t)-\mathcal{C}_{1}\left(
\rho _{1}^{(N)}|\rho _{1}^{(N)}\right) =0, \\
\rho _{1}^{(N)}(\mathbf{x}_{1},t_{o})=\rho _{1o}^{(N)}(\mathbf{x}_{1}),%
\end{array}%
\right.  \label{App-1}
\end{equation}%
\ can be shown to admit a unique global solution \cite{noi7}. Here the
notation is standard \cite{noi3}. Thus%
\begin{eqnarray}
&&\left. \mathcal{C}_{1}\left( \rho _{1}^{(N)}|\rho _{1}^{(N)}\right) \equiv
\left( N-1\right) \sigma ^{2}\int\limits_{U_{1(2)}}d\mathbf{v}%
_{2}\int^{(-)}d\Sigma _{21}\right.  \notag \\
&&\left[ \widehat{\rho }_{1}^{(N)}(\mathbf{r}_{1},\mathbf{v}_{1}^{(+)},t)%
\widehat{\rho }_{1}^{(N)}(\mathbf{r}_{2},\mathbf{v}_{2}^{(+)},t)-\widehat{%
\rho }_{1}^{(N)}(\mathbf{r}_{1},\mathbf{v}_{1},t)\widehat{\rho }_{1}^{(N)}(%
\mathbf{r}_{2},\mathbf{v}_{2},t)\right] \times  \notag \\
&&\left\vert \mathbf{v}_{21}\cdot \mathbf{n}_{21}\right\vert k_{2}^{(N)}(%
\mathbf{r}_{1},\mathbf{r}_{2},t)\overline{\Theta }^{\ast }(\mathbf{r}_{2})
\label{App-2}
\end{eqnarray}%
identifies the Master collision operator while $\rho _{1o}^{(N)}(\mathbf{x}%
_{1})$ is the initial $1-$body PDF which belongs to the functional class $%
\left\{ \rho _{1o}^{(N)}(\mathbf{x}_{1})\right\} $ of stochastic, i.e.,
strictly-positive, smooth ordinary functions, $1-$body PDFs. Furthermore,
the solid-angle integral on the rhs of Eq.(\ref{App-2}) is now evaluated on
the subset in which $\mathbf{v}_{12}\cdot \mathbf{n}_{12}<0,$ while $\mathbf{%
r}_{2}$ identifies\emph{\ }$\mathbf{r}_{2}=\mathbf{r}_{1}+\sigma \mathbf{n}%
_{21}$, while $k_{1}^{(N)}(\mathbf{r}_{1},t)$ and $k_{2}^{(N)}(\mathbf{r}%
_{1},\mathbf{r}_{2},t)$ coincide respectively with the $1-$ and $2-$body
occupation coefficients \cite{noi3} and $\overline{\Theta }^{\ast }\equiv
\overline{\Theta }^{\ast }(\mathbf{r}_{i})$ is prescribed by$\ $%
\begin{equation}
\overline{\Theta }^{\ast }(\mathbf{r}_{i})\equiv \overline{\Theta }%
_{i}^{(\partial \Omega )}(\overline{\mathbf{r}})\equiv \overline{\Theta }%
\left( \left\vert \mathbf{r}_{i}-\frac{\sigma }{2}\mathbf{n}_{i}\right\vert -%
\frac{\sigma }{2}\right)  \label{App-2b}
\end{equation}%
with $\overline{\Theta }(x)$ being the strong Heaviside theta function $%
\overline{\Theta }(x)=\left\{
\begin{array}{lll}
1 &  & y>0 \\
0 &  & y\leq 0%
\end{array}%
\right. $.

Regarding the specific identification of the occupation coefficients let us
preliminarily recall the notion of $S_{N}-$\emph{\ ensemble strong
theta-function} $\overline{\Theta }^{(N)}.$ The latter is prescribed,
according to Ref.\cite{noi3}, by requiring that
\begin{equation}
\overline{\Theta }^{(N)}(\overline{\mathbf{r}})=1  \label{CONSTRAINT-1}
\end{equation}%
for all configuration vectors $\overline{\mathbf{r}}\equiv \left\{ \mathbf{r}%
_{1},...,\mathbf{r}_{N}\right\} $ belonging to the collisionless subset of $%
\Omega ^{(N)}$. This is identified with the open subset of the $N-$body
configuration domain $\Omega ^{(N)}\equiv \prod\limits_{i=1,N}\Omega $ in
which each of the particles of $S_{N}$ is not in mutual contact with any
other particle of $S_{N}$ or with the boundary $\theta \Omega $ of $\Omega .$
this can be prescribed in terms of the \emph{n,} \textit{i.e.}, in such a
way that identically In agreement with Ref.\cite{noi3} this occurs for $N-$%
body PDFs which are represented by ordinary functions (i.e., are
stochastic). This means that $\overline{\Theta }^{(N)}(\overline{\mathbf{r}}%
) $ can be prescribed as

\begin{equation}
\overline{\Theta }^{(N)}(\overline{\mathbf{r}})\equiv \prod\limits_{i=1,N}%
\overline{\Theta }_{i}(\overline{\mathbf{r}})\overline{\Theta }%
_{i}^{(\partial \Omega )}(\overline{\mathbf{r}}).  \label{ENSEMBLE-THETA-1}
\end{equation}%
Here $\overline{\Theta }_{i}^{(\partial \Omega )}(\overline{\mathbf{r}})$
identifies the $i-$th particle "boundary" theta function%
\begin{equation}
\overline{\Theta }_{i}^{(\partial \Omega )}(\overline{\mathbf{r}})\equiv
\overline{\Theta }\left( \left\vert \mathbf{r}_{i}-\mathbf{r}%
_{Wi}\right\vert -\frac{\sigma }{2}\right) ,  \label{boundary theta function}
\end{equation}%
with $\mathbf{r}_{Wi}=\mathbf{r}_{i}-\rho \mathbf{n}_{i}$ and $\rho \mathbf{n%
}_{i}$ the inward vector normal to the boundary belonging to the center of
the $i-$th particle having a distance $\rho $ from the same boundary.
Furthermore $\overline{\Theta }_{i}(\overline{\mathbf{r}})$ is the
"binary-collision" theta function. A possible identification of $\overline{%
\Theta }_{i}(\overline{\mathbf{r}})$ which warrants validity of Eq.(\ref%
{CONSTRAINT-1}) is manifestly given by the expression
\begin{equation}
\overline{\Theta }_{i}(\overline{\mathbf{r}})\equiv \prod\limits_{\substack{ %
j=1,N;  \\ i<j}}\overline{\Theta }\left( \left\vert \mathbf{r}_{i}-\mathbf{r}%
_{j}\right\vert -\sigma \right) ,  \label{ENSEMBLE.THETA-2A}
\end{equation}%
namely%
\begin{equation}
\left\{
\begin{array}{c}
\overline{\Theta }_{i}(\overline{\mathbf{r}})\equiv \prod\limits_{\substack{ %
j=1,N;  \\ i<j}}\overline{\Theta }_{ij}(\overline{\mathbf{r}}), \\
\overline{\Theta }_{ij}(\overline{\mathbf{r}})\equiv \overline{\Theta }%
\left( \left\vert \mathbf{r}_{i}-\mathbf{r}_{j}\right\vert -\sigma \right) .%
\end{array}%
\right.  \label{ENSEMBLE THETA-2A-FACTOR}
\end{equation}%
However an equivalent possible prescription of $\overline{\Theta }_{i}(%
\overline{\mathbf{r}})$ is also provided by the alternative realization
obtained letting%
\begin{equation}
\left\{
\begin{array}{c}
\overline{\Theta }_{i}(\overline{\mathbf{r}})\equiv \prod\limits_{\substack{ %
j=1,N;  \\ i<j}}\prod\limits_{\substack{ m,n=1,N  \\ i<m<n}}\overline{\Theta
}_{ij}^{mn}(\overline{\mathbf{r}}), \\
\overline{\Theta }_{ij}^{mn}(\overline{\mathbf{r}})\equiv \overline{\Theta }%
\left( \left\vert \mathbf{r}_{i}-\mathbf{r}_{j}\right\vert -\sigma \right)
\times \\
\overline{\Theta }(\left\vert \mathbf{r}_{i}-\mathbf{r}_{m}\right\vert
+\left\vert \mathbf{r}_{i}-\mathbf{r}_{n}\right\vert -2\sigma )\overline{%
\Theta }(\left\vert \mathbf{r}_{m}-\mathbf{r}_{n}\right\vert +\left\vert
\mathbf{r}_{i}-\mathbf{r}_{m}\right\vert -2\sigma ).%
\end{array}%
\right.  \label{Factro-ENSEMBLE-THETA-2}
\end{equation}%
\bigskip Indeed by construction in the subset of $\Omega ^{(N)}$ in which
for all $i=1,N$ the rhs of Eq.(\ref{ENSEMBLE.THETA-2A}) is identically equal
to unity the factor $\prod\limits_{\substack{ m,n=1,N  \\ i<m<n}}\overline{%
\Theta }(\left\vert \mathbf{r}_{i}-\mathbf{r}_{m}\right\vert +\left\vert
\mathbf{r}_{i}-\mathbf{r}_{n}\right\vert -2\sigma )\overline{\Theta }%
(\left\vert \mathbf{r}_{m}-\mathbf{r}_{n}\right\vert +\left\vert \mathbf{r}%
_{i}-\mathbf{r}_{m}\right\vert -2\sigma )$ is necessarily equal to unity
too. Incidentally, we notice in fact that the latter factor carries the
contributions due to triple collisions which are by construction ruled out
in the domain of validity of Eq.(\ref{CONSTRAINT-1}).

\section{Appendix B - Integral and differential identities}

One notices that although the definitions (\ref{Factro-ENSEMBLE-THETA-2})
and (\ref{ENSEMBLE.THETA-2A}) given in Appendix A for $\overline{\Theta }%
_{i}(\overline{\mathbf{r}})$ coincide in the collisionless subset of $\Omega
^{(N)},$ only the first one is applicable in the complementary collision
subset. Based on these premises in this appendix a number of integral and
differential identities holding for the $1-$ and $2-$body occupation
coefficients are displayed.

First, recalling Ref.\cite{noi3}, one notices that the realizations of the $%
1-$ and $s-$body occupation coefficients $k_{1}^{(N)}(\mathbf{r}%
_{i},t),k_{2}^{(N)}(\mathbf{r}_{1},\mathbf{r}_{2},t),$..., $k_{s}^{(N)}(%
\mathbf{r}_{1},\mathbf{r}_{2},..\mathbf{r}_{s},t)$ remain uniquely
prescribed by the $1-$body PDF, being given by%
\begin{eqnarray}
k_{1}^{(N)}(\mathbf{r}_{1},t) &\equiv &F_{1}\left\{ \prod\limits_{j=2,N}%
\frac{\rho _{1}^{(N)}(\mathbf{x}_{j},t)}{k_{1}^{(N)}(\mathbf{r}_{j},t)}%
\right\} ,  \label{App-4} \\
k_{2}^{(N)}(\mathbf{r}_{1},\mathbf{r}_{2},t) &\equiv &F_{2}\left\{
\prod\limits_{j=s+1,N}\frac{\rho _{1}^{(N)}(\mathbf{x}_{j},t)}{k_{1}^{(N)}(%
\mathbf{r}_{j},t)}\right\} ,  \label{App-5} \\
&&...  \notag \\
k_{s}^{(N)}(\mathbf{r}_{1},\mathbf{r}_{2},..\mathbf{r}_{s},t) &\equiv
&F_{s}\left\{ \prod\limits_{j=s+1,N}\frac{\rho _{1}^{(N)}(\mathbf{x}_{j},t)}{%
k_{1}^{(N)}(\mathbf{r}_{j},t)}\right\} ,  \label{App-6a}
\end{eqnarray}%
\textbf{\ }where $F_{s}$\ denotes the integral operator%
\begin{equation}
F_{s}\equiv \int\limits_{\Gamma _{N}}d\overline{\mathbf{x}}\overline{\Theta }%
^{(N)}(\overline{\mathbf{r}})\prod\limits_{i=1,s}\delta (\mathbf{x}_{i}-%
\overline{\mathbf{x}}_{i}).  \label{App-6}
\end{equation}%
Therefore, since in the collisionless subset of $\Omega ^{(N)}$ the
prescriptions (\ref{ENSEMBLE.THETA-2A}) and (\ref{Factro-ENSEMBLE-THETA-2})
are equivalent, in the same subset the $1-$\ and $2-$body occupation
coefficients, written in terms of Eq.(\ref{ENSEMBLE.THETA-2A}), become
explicitly%
\begin{eqnarray}
&&\left. k_{1}^{(N)}(\mathbf{r}_{1},t)=\int\limits_{\Gamma _{1(2)}}d\mathbf{x%
}_{2}\frac{\rho _{1}^{(N)}(\mathbf{x}_{2},t)}{k_{1}^{(N)}(\mathbf{r}_{2},t)}%
\Theta _{2}^{(\partial \Omega )}(\overline{\mathbf{r}})\overline{\Theta }%
\left( \left\vert \mathbf{r}_{2}-\mathbf{r}_{1}\right\vert -\sigma \right)
\times \right.  \notag \\
&&\int\limits_{\Gamma _{1(3)}}d\mathbf{x}_{3}\frac{\rho _{1}^{(N)}(\mathbf{x}%
_{3},t)}{k_{1}^{(N)}(\mathbf{r}_{3},t)}\Theta _{3}^{(\partial \Omega )}(%
\overline{\mathbf{r}})\prod\limits_{j=1,2}\overline{\Theta }\left(
\left\vert \mathbf{r}_{3}-\mathbf{r}_{j}\right\vert -\sigma \right) ....
\notag \\
&&\left. ...\int\limits_{\Gamma _{1(N)}}d\mathbf{x}_{N}\frac{\rho _{1}^{(N)}(%
\mathbf{x}_{N},t)}{k_{1}^{(N)}(\mathbf{r}_{N},t)}\Theta _{N}^{(\partial
\Omega )}(\overline{\mathbf{r}})\prod\limits_{j=1,N-1}\overline{\Theta }%
\left( \left\vert \mathbf{r}_{N}-\mathbf{r}_{j}\right\vert -\sigma \right)
,\right.  \label{App-7a}
\end{eqnarray}%
and%
\begin{eqnarray}
&&\left. k_{2}^{(N)}(\mathbf{r}_{1},\mathbf{r}_{2},t)=\int\limits_{\Gamma
_{1(3)}}d\mathbf{x}_{3}\frac{\rho _{1}^{(N)}(\mathbf{x}_{3},t)}{k_{1}^{(N)}(%
\mathbf{r}_{3},t)}\overline{\Theta }_{3}^{(\partial \Omega )}(\overline{%
\mathbf{r}})\prod\limits_{j=1,2}\overline{\Theta }\left( \left\vert \mathbf{r%
}_{3}-\mathbf{r}_{j}\right\vert -\sigma \right) \times \right.  \notag \\
&&\int\limits_{\Gamma _{1(4)}}d\mathbf{x}_{4}\frac{\rho _{1}^{(N)}(\mathbf{x}%
_{4},t)}{k_{1}^{(N)}(\mathbf{r}_{4},t)}\Theta _{4}^{(\partial \Omega )}(%
\overline{\mathbf{r}})\prod\limits_{j=1,3}\overline{\Theta }\left(
\left\vert \mathbf{r}_{4}-\mathbf{r}_{j}\right\vert -\sigma \right) ... \\
&&....\int\limits_{\Gamma _{1(N)}}d\mathbf{x}_{N}\frac{\rho _{1}^{(N)}(%
\mathbf{x}_{N},t)}{k_{1}^{(N)}(\mathbf{r}_{N},t)}\Theta _{N}^{(\partial
\Omega )}(\overline{\mathbf{r}})\prod\limits_{j=1,N-1}\overline{\Theta }%
\left( \left\vert \mathbf{r}_{N}-\mathbf{r}_{j}\right\vert -\sigma \right) .
\label{App-8a}
\end{eqnarray}%
Accordingly letting $\mathbf{n}_{jj}=\mathbf{r}_{uij}/\left\vert \mathbf{r}%
_{ij}\right\vert $ with $\mathbf{r}_{ij}=\mathbf{r}_{i}-\mathbf{r}_{j}$, one
notices that in the collisionless subset of $\Omega ^{(N)}$ the following
differential identities hold for all $s=1,N-1$:

\begin{eqnarray}
&&\frac{\partial }{\partial \mathbf{r}_{1}}k_{1}^{(N)}(\mathbf{r}%
_{1},t)=\left( N-1\right) \int\limits_{\Gamma _{1(2)}}d\mathbf{x}_{2}\mathbf{%
n}_{12}\delta \left( \left\vert \mathbf{r}_{2}-\mathbf{r}_{1}\right\vert
-\sigma \right) \times  \notag \\
&&k_{2}^{(N)}(\mathbf{r}_{1},\mathbf{r}_{2},t)\overline{\Theta }%
_{2}^{(\partial \Omega )}(\overline{\mathbf{r}})\widehat{\rho }_{1}^{(N)}(%
\mathbf{x}_{2},t),  \label{App-X1} \\
&&\left\{
\begin{array}{c}
\frac{\partial }{\partial \mathbf{r}_{1}}k_{2}^{(N)}(\mathbf{r}_{1},\mathbf{r%
}_{2},t)=\left( N-2\right) \int\limits_{\Gamma _{1(2)}}d\mathbf{x}_{3}%
\mathbf{n}_{13}\delta \left( \left\vert \mathbf{r}_{3}-\mathbf{r}%
_{1}\right\vert -\sigma \right) \times \\
\prod\limits_{j=1,2;j\neq 1}\overline{\Theta }\left( \left\vert \mathbf{r}%
_{3}-\mathbf{r}_{j}\right\vert -\sigma \right) \overline{\Theta }%
_{3}^{(\partial \Omega )}(\overline{\mathbf{r}})k_{3}^{(N)}(\mathbf{r}_{1},%
\mathbf{r}_{2},\mathbf{r}_{3},t)\widehat{\rho }_{1}^{(N)}(\mathbf{x}_{3},t)
\\
\frac{\partial }{\partial \mathbf{r}_{2}}k_{2}^{(N)}(\mathbf{r}_{1},\mathbf{r%
}_{2},t)=\left( N-2\right) \int\limits_{\Gamma _{1(2)}}d\mathbf{x}_{3}%
\mathbf{n}_{23}\delta \left( \left\vert \mathbf{r}_{3}-\mathbf{r}%
_{2}\right\vert -\sigma \right) \times \\
\prod\limits_{j=1,2;j\neq 2}\overline{\Theta }\left( \left\vert \mathbf{r}%
_{3}-\mathbf{r}_{j}\right\vert -\sigma \right) \overline{\Theta }%
_{3}^{(\partial \Omega )}(\overline{\mathbf{r}})k_{3}^{(N)}(\mathbf{r}_{1},%
\mathbf{r}_{2},\mathbf{r}_{3},t)\widehat{\rho }_{1}^{(N)}(\mathbf{x}_{3},t)%
\end{array}%
\right.  \label{App-X2} \\
&&...............  \notag \\
&&\left\{
\begin{array}{c}
\frac{\partial }{\partial \mathbf{r}_{1}}k_{s}^{(N)}(\mathbf{r}_{1},\mathbf{r%
}_{2},..,\mathbf{r}_{s},t)=\left( N-s\right) \int\limits_{\Gamma _{1(2)}}d%
\mathbf{x}_{s+1}\mathbf{n}_{1s+1}\delta \left( \left\vert \mathbf{r}_{s+1}-%
\mathbf{r}_{1}\right\vert -\sigma \right) \times \\
\prod\limits_{j=1,s;j\neq 1}\overline{\Theta }\left( \left\vert \mathbf{r}%
_{s+1}-\mathbf{r}_{j}\right\vert -\sigma \right) \overline{\Theta }%
_{3}^{(\partial \Omega )}(\overline{\mathbf{r}})k_{s+1}^{(N)}(\mathbf{r}_{1},%
\mathbf{r}_{2},.,\mathbf{r}_{s+1},t)\widehat{\rho }_{1}^{(N)}(\mathbf{x}%
_{s+1},t) \\
\frac{\partial }{\partial \mathbf{r}_{2}}k_{s}^{(N)}(\mathbf{r}_{1},\mathbf{r%
}_{2},..,\mathbf{r}_{s},t)=\left( N-s\right) \int\limits_{\Gamma _{1(2)}}d%
\mathbf{x}_{s+1}\mathbf{n}_{2s+1}\delta \left( \left\vert \mathbf{r}_{s+1}-%
\mathbf{r}_{2}\right\vert -\sigma \right) \times \\
\prod\limits_{j=1,s;j\neq 2}\overline{\Theta }\left( \left\vert \mathbf{r}%
_{s+1}-\mathbf{r}_{j}\right\vert -\sigma \right) \overline{\Theta }%
_{3}^{(\partial \Omega )}(\overline{\mathbf{r}})k_{s+1}^{(N)}(\mathbf{r}_{1},%
\mathbf{r}_{2},.,\mathbf{r}_{s+1},t)\widehat{\rho }_{1}^{(N)}(\mathbf{x}%
_{s+1},t) \\
..... \\
\frac{\partial }{\partial \mathbf{r}_{s}}k_{s}^{(N)}(\mathbf{r}_{1},\mathbf{r%
}_{2},..,\mathbf{r}_{s},t)=\left( N-s\right) \int\limits_{\Gamma _{1(2)}}d%
\mathbf{x}_{s+1}\mathbf{n}_{ss+1}\delta \left( \left\vert \mathbf{r}_{s+1}-%
\mathbf{r}_{s}\right\vert -\sigma \right) \times \\
\prod\limits_{j=1,s;j\neq s}\overline{\Theta }\left( \left\vert \mathbf{r}%
_{s+1}-\mathbf{r}_{j}\right\vert -\sigma \right) \overline{\Theta }%
_{3}^{(\partial \Omega )}(\overline{\mathbf{r}})k_{s+1}^{(N)}(\mathbf{r}_{1},%
\mathbf{r}_{2},.,\mathbf{r}_{s+1},t)\widehat{\rho }_{1}^{(N)}(\mathbf{x}%
_{s+1},t)%
\end{array}%
\right.  \label{App-X3}
\end{eqnarray}%
As a consequence the following identities (the first one needed to evaluate
the rhs of Eq.(\ref{APP-0b}) in Appendix A)
\begin{eqnarray}
L_{1\left( 1\right) }k_{1}^{(N)}(\mathbf{r}_{1},t) &=&\left( N-1\right)
\int\limits_{\Gamma _{1(2)}}d\mathbf{x}_{2}\mathbf{v}_{21}\cdot \mathbf{n}%
_{21}\delta \left( \left\vert \mathbf{r}_{2}-\mathbf{r}_{1}\right\vert
-\sigma \right)  \notag \\
&&\overline{\Theta }^{\ast }(\mathbf{r}_{2})k_{2}^{(N)}(\mathbf{r}_{1},%
\mathbf{r}_{2},t)\widehat{\rho }_{1}^{(N)}(\mathbf{x}_{2},t),
\label{DIFF-identity}
\end{eqnarray}%
\begin{eqnarray}
&&\left. \frac{\partial ^{2}k_{1}^{(N)}(\mathbf{r}_{1},t)}{\partial \mathbf{r%
}_{1}\cdot \partial \mathbf{r}_{1}}=-\left( N-1\right) \int\limits_{\Gamma
_{1(2)}}d\mathbf{x}_{2}k_{2}^{(N)}(\mathbf{r}_{1},\mathbf{r}_{2},t)\delta
\left( \left\vert \mathbf{r}_{2}-\mathbf{r}_{1}\right\vert -\sigma \right)
\right.  \notag \\
&&\overline{\Theta }_{2}^{(\partial \Omega )}(\overline{\mathbf{r}})\mathbf{n%
}_{21}\cdot \frac{\partial }{\partial \mathbf{r}_{2}}\widehat{\rho }%
_{1}^{(N)}(\mathbf{x}_{2},t),  \label{Diff-identity-2}
\end{eqnarray}%
hold too. However, the alternative realization of the factor $\overline{%
\Theta }_{i}(\overline{\mathbf{r}})$ given by Eq.(\ref%
{Factro-ENSEMBLE-THETA-2}) (see Appendix A) has the virtue of excluding
explicitly explicitly multiple collisions. The consequence is that when such
a definition is adopted the differential identities%
\begin{equation}
\left\{
\begin{array}{c}
\delta \left( \left\vert \mathbf{r}_{2}-\mathbf{r}_{1}\right\vert -\sigma
\right) \frac{\partial }{\partial \mathbf{r}_{1}}k_{2}^{(N)}(\mathbf{r}_{1},%
\mathbf{r}_{2},t)=0, \\
\delta \left( \left\vert \mathbf{r}_{2}-\mathbf{r}_{1}\right\vert -\sigma
\right) \frac{\partial }{\partial \mathbf{r}_{2}}k_{2}^{(N)}(\mathbf{r}_{1},%
\mathbf{r}_{2},t)=0,%
\end{array}%
\right.  \label{App-10a}
\end{equation}%
both hold identically. The latter equations, in fact, manifestly hold also
in the collision subset where $\delta \left( \left\vert \mathbf{r}_{2}-%
\mathbf{r}_{1}\right\vert -\sigma \right) \neq 0.$

\section{Appendix C: Causal and anti-causal forms of collisional boundary
conditions}

For definiteness, let us denote respectively the outgoing and incoming $N-$%
body PDFs $\rho ^{(-)\left( N\right) }(\mathbf{x}^{\left( -\right)
}(t_{i}),t_{i})$ and $\rho ^{(+)\left( N\right) }(\mathbf{x}^{\left(
+\right) }(t_{i}),t_{i})$, with $\rho ^{(\pm )\left( N\right) }(\mathbf{x}%
^{\left( \pm \right) }(t_{i}),t_{i})=\lim_{t\rightarrow t_{i}^{\left( \pm
\right) }}\rho ^{\left( N\right) }(\mathbf{x}(t),t)$, where $\mathbf{x}%
^{(-)}(t_{i})$ and $\mathbf{x}^{(+)}(t_{i})$, with $\mathbf{x}^{\left( \pm
\right) }(t_{i})=\lim_{t\rightarrow t_{i}^{\left( \pm \right) }}\mathbf{x}%
(t),$ are the incoming and outgoing Lagrangian $N-$body states, their mutual
relationship being again determined by the collision laws holding for the $%
S_{N}-$CDS. Here it is understood that:

\begin{itemize}
\item The $S_{N}-$CDS is referred to a reference frame $O\left( \mathbf{r}%
,\tau \equiv t-t_{o}\right) $, having respectively spatial and time origins
at the point $O$ which belongs to the Euclidean space $%
%TCIMACRO{\U{211d} }%
%BeginExpansion
\mathbb{R}
%EndExpansion
^{3}$ and at time $t_{o}\in I$.

\item In addition, by assumption the time-axis is oriented. Such an
orientation is referred to as \emph{microscopic arrow of time}.
\end{itemize}

For an arbitrary $N-$body PDF $\rho ^{\left( N\right) }(\mathbf{x},t)$
belonging to the extended functional setting and an arbitrary collision
event occurring at time $t_{i}$ two possible realizations of the MCBC can in
principle be given, both yielding a relationship between the PDFs $\rho
^{(+)\left( N\right) }$ and $\rho ^{(-)\left( N\right) }$. In the context of
the "\textit{ab initio}" statistical approach\ based on the Master kinetic
equation \cite{noi1,noi2,noi3,noi4,noi5,noi6,noi7} these are provided by the
two possible realizations of the so-called \emph{modified CBC} (MCBC). When
expressed in Lagrangian form they are realized respectively either by the
\emph{causal} and\emph{\ anti-causal MCBC}, namely%
\begin{equation}
\rho ^{(+)\left( N\right) }(\mathbf{x}^{\left( +\right) }(t_{i}),t_{i})=\rho
^{(-)\left( N\right) }(\mathbf{x}^{(+)}(t_{i}),t_{i}),  \label{bbb1}
\end{equation}%
or%
\begin{equation}
\rho ^{(-)\left( N\right) }(\mathbf{x}^{\left( -\right) }(t_{i}),t_{i})=\rho
^{(+)\left( N\right) }(\mathbf{x}^{(-)}(t_{i}),t_{i}).  \label{bbb2}
\end{equation}%
The corresponding Eulerian forms of the MCBC can easily be determined (see
Ref.\cite{noi6}). The one corresponding to Eq.(\ref{bbb1}) is, for example,
provided\ by the condition%
\begin{equation}
\rho ^{(+)\left( N\right) }(\mathbf{x}^{\left( +\right) },t)=\rho
^{(-)\left( N\right) }(\mathbf{x}^{(+)},t),  \label{bbb3}
\end{equation}%
where now $\mathbf{x}^{\left( +\right) }$ denotes again an arbitrary
outgoing collision state.

Once the time-axis is oriented, \textit{i.e.}, the \emph{microscopic} \emph{%
arrow of time} is prescribed, the validity of the causality principle in the
reference frame $\left( \mathbf{r},\tau \equiv t-t_{o}\right) $ manifestly
requires invoking Eq.(\ref{bbb1}). Indeed, Eq.(\ref{bbb1}) predicts the
future (\textit{i.e.}, outgoing) PDF from the past (incoming) one. Therefore
the choice (\ref{bbb1}) is the one which is manifestly consistent with the
causality principle. On the other hand, if the arrow of time is changed,
\textit{i.e.} the time-reversal transformation with respect to the initial
time (or time-origin) $t_{o},$ \textit{i.e.}, the map between the two
reference frames%
\begin{equation}
O\left( \mathbf{r},\tau \equiv t-t_{o}\right) \rightarrow O\left( \mathbf{r}%
,\tau ^{\prime }\right) ,
\end{equation}%
with $\tau ^{\prime }=-\tau $ is performed, it is obvious that for the
transformed\ reference frame $O\left( \mathbf{r},\tau ^{\prime }\right) $
the form of CBC consistent with causality principle becomes that given by
Eq.(\ref{bbb2}). Analogous conclusions hold if a velocity-reversal is
performed, implying the incoming states and corresponding PDF must be
exchanged with corresponding outgoing ones and vice versa.

\section{Appendix D: Treatment of case $N=2$}

For completeness let us briefly comment on the particular realization of
MPI/DKE theory which is achieved in the special case $N=2.$\ For this
purpose, one notices that - thanks to Eq.(\ref{App-5}) recalled in Appendix
B (see also Ref.\cite{noi3}) - in this case by construction $k_{2}^{(N)}(%
\mathbf{r}_{1},\mathbf{r}_{2},t)$\ simply reduces to%
\begin{equation}
k_{2}^{(N)}(\mathbf{r}_{1},\mathbf{r}_{2},t)\equiv 1.  \label{AppD-1}
\end{equation}%
Accordingly, once the same prescription is invoked, both the Master kinetic
equation (\ref{App-1}) and the corresponding Master collision operator (\ref%
{App-2}) remain formally unchanged. In a similar way it is important to
remark that the expression of the functional $\frac{\partial }{\partial t}%
K_{M}(\rho _{1}^{(N)}(t),\mathbf{b})\equiv W_{M}(\rho _{1}^{(N)}(t),\mathbf{b%
})$ given by Eq. (\ref{P2-2-2d}) is still correct also in such a case, being
now given by
\begin{eqnarray}
&&\left. \frac{\partial }{\partial t}K_{M}(\rho _{1}^{(N)}(t),\mathbf{b}%
)\equiv W_{M}(\rho _{1}^{(N)}(t),\mathbf{b})=-(N-1)\sigma
^{2}\int\limits_{U_{1(1)}}d\mathbf{v}_{1}\int\limits_{U_{1(2)}}d\mathbf{v}%
_{2}\times \right.  \notag \\
&&\int\limits_{\Omega }d\mathbf{r}_{1}\int^{(-)}d\mathbf{\Sigma }_{21}\frac{%
\partial \widehat{\rho }_{1}^{(N)}(\mathbf{r}_{1},\mathbf{v}_{1}^{(+)},t)}{%
\partial \mathbf{r}_{1}}\cdot \frac{\partial \widehat{\rho }_{1}^{(N)}(%
\mathbf{r}_{2}=\mathbf{r}_{1}+\sigma \mathbf{n}_{21},\mathbf{v}_{2}^{(+)}t)}{%
\partial \mathbf{r}_{2}}\times  \notag \\
&&\left. \left\vert \mathbf{v}_{12}^{(+)}\cdot \mathbf{n}_{12}\right\vert
\left( \mathbf{b\cdot n}_{12}\right) ^{2}\left( \mathbf{n}_{12}\cdot \mathbf{%
v}_{12}^{(+)}\right) ^{2}\leq 0.\right.  \label{AAPD-2}
\end{eqnarray}

It is then immediate to infer the validity of both the PMI theorem (THM.2)
and the DKE property for the Master kinetic equation (THM.3). As a
consequence one concludes that MPI/DKE theory holds also in the special case
$N=2,$ This conclusion is not unexpected. In fact, binary collisions, as
indicated above, are responsible for the MPI/DKE phenomenology and in such a
case can only occur between particles $1$\ and $2.$

\end{document}